\documentclass[journal]{IEEEtran}%
\usepackage{amsfonts}
\usepackage{amsmath}
\usepackage{amssymb}
\usepackage{graphicx}%
\newtheorem{theorem}{Theorem}

\newtheorem{definition}[theorem]{Definition}
\newtheorem{example}[theorem]{Example}

\newtheorem{lemma}[theorem]{Lemma}

\newtheorem{proposition}[theorem]{Proposition}
\newtheorem{remark}[theorem]{Remark}

\graphicspath{{c:/LS_laptop_phdwork/phd_work/JSAC_GameTheory/}}
\begin{document}

\title{Coalitions in Cooperative Wireless Networks}
\author{Suhas Mathur,~\IEEEmembership{Member,~IEEE,} Lalitha
Sankar,~\IEEEmembership{Member,~IEEE,} and~Narayan B.
Mandayam,~\IEEEmembership{Senior~Member,~IEEE}\thanks{Manuscript received
August 15, 2007; revised January 28, 2008. }\thanks{The work of S. Mathur,
L.~Sankar (previously Sankaranarayanan) and N.~B.~Mandayam was supported in
part by the National\ Science Foundation under
Grant~No.~{\scriptsize TF:0634973}. . The material in this paper was presented
in part at the IEEE International Symposium on Information Theory, Seattle,
WA, Jul. 2006; at the IEEE\ Conference on Information Sciences and Systems,
Princeton, NJ, Mar. 2006; at the $40^{th}$ Annual Asilomar Conference
on\ Signals, Systems, and Computers, Pacific Grove, CA, Nov. 2006; and at the
Information Theory and Applications Workshop, San Diego, CA, Jan. 2008. S.
Mathur and N. B. Mandayam are with the WINLAB, Department of Electrical
Engineering, Rutgers University, Technology Center of NJ, 671 Route 1S, North
Brunswick, NJ, 08902. Email: suhas@winlab.rutgers.edu and
narayan@winlab.rutgers.edu. L. Sankar is with the Department of Electrical
Engineering, Princeton University, E-Quad, Olden Street, Princeton, NJ 08854.
Email: lalitha@princeton.edu.}}
\pubid{~}
\specialpapernotice{~}
\maketitle

\begin{abstract}
Cooperation between rational users in wireless networks is studied using
coalitional game theory. Using the rate achieved by a user as its utility, it
is shown that the stable coalition structure, i.e., set of coalitions from
which users have no incentives to defect, depends on the manner in which the
rate gains are apportioned among the cooperating users. Specifically, the
stability of the \textit{grand coalition} (GC), i.e., the coalition of all
users, is studied. Transmitter and receiver cooperation in an interference
channel (IC) are studied as illustrative cooperative models to determine the
stable coalitions for both flexible (\textit{transferable}) and fixed
(\textit{non-transferable}) apportioning schemes. It is shown that the stable
sum-rate optimal coalition when only receivers cooperate by jointly decoding
(transferable) is the GC. The stability of the GC depends on the detector when
receivers cooperate using linear multiuser detectors (non-transferable).
Transmitter cooperation is studied assuming that all receivers cooperate
perfectly and that users outside a coalition act as jammers. The stability of
the GC is studied for both the case of perfectly cooperating transmitters
(transferrable) and under a \textit{partial decode-and-forward} strategy
(non-transferable). In both cases, the stability is shown to depend on the
channel gains and the transmitter jamming strengths.

\end{abstract}

\begin{keywords}
Coalitional games, cooperative communications, interference channel.
\end{keywords}

\section{Introduction}

Cooperation in wireless networks results when nodes exploit the broadcast
nature of the wireless medium and use their power and bandwidth resources to
mutually enhance transmissions (see, for e.g., \cite{SEA:01,
cap_theorems:LWT01, cap_theorems:KGG_IT} and the references therein). In
general, it is assumed that all the network nodes are willing to cooperate.
However, when rational (self-interested) users are allowed to cooperate it is
necessary to examine whether the cooperation of all users, i.e., the
\textit{grand coalition} (GC) of all users, can be taken for granted. In fact,
cooperation may involve significant costs and the greatest immediate benefits
may not be achieved by the users that bear the greatest immediate cost. An
additional disincentive to cooperation may result from the rules by which the
cooperative gains are distributed among participating users. In fact, for
maximum gains users may prefer to cooperate with a select set of users to form
\textit{coalitions} that are closed to cooperation from users outside the
group. For example, consider a multi-user wireless network where users labeled
$A$, $B$, and $C$ are decoded at a central receiver. Cooperating users share
the benefit of having their signals jointly decoded at the receiver while a
user that chooses not to cooperate is decoded independently and is subject to
interference from the other users.

\begin{figure}
[ptb]
\begin{center}
\includegraphics[
height=2.5529in,
width=3.3416in
]%
{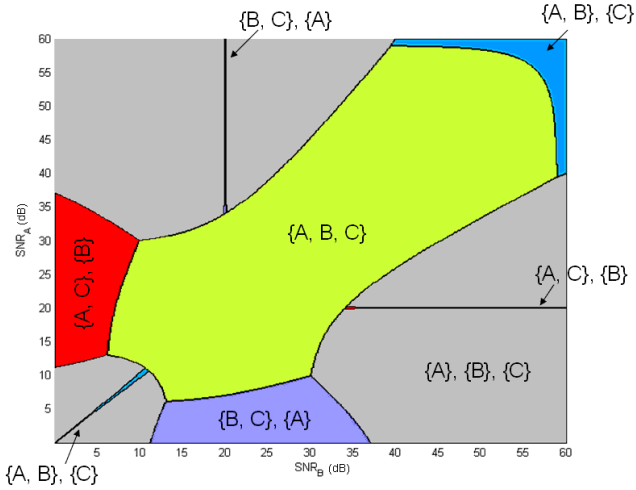}%
\caption{Stable Coalition Structures as a function of the SNR values of users
$A$ and $B$ and $SNR_{C}$ $=20$ dB.}%
\label{Fig_ABCStructure}%
\end{center}
\end{figure}

One can verify that the multiaccess channel (MAC) that results when all three
users cooperate achieves the maximum information-theoretic three-user sum-rate
\cite[14.3]{cap_theorems:CTbook}. However, it is not clear if the GC is also a
\textit{stable} coalition, i.e., a coalition whose users do not have an
incentive to leave (for larger rates). For example, consider an apportionment
strategy where the sum-rate achieved is divided equally among the users in a
coalition. In Fig. \ref{Fig_ABCStructure} we demonstrate the stability of the
various coalitions as a function of the received signal-to-noise ratio (SNR)
of each user. Observe that the grand coalition is desirable only when all
users have similar SNR values. Further, for arbitrary SNR values, the users in
the stable coalitions benefit from the exclusion of the weak interferer. Thus,
even in this relatively simple example we see that user cooperation is
desirable only when the aggregate benefits of cooperation provide adequate
incentives to all participating users.

We use the framework of coalitional game theory to determine the stable
coalition structure, i.e., a set of coalitions whose users do not have
incentives to break away, when wireless nodes are allowed to cooperate (see
for e.g., \cite{LA:01,Rodoplu_Meng}). We consider a $K$-link interference
channel (IC)\ \cite{ABC:IntCh} as an illustrative network model to determine
the stable coalitions when transmitters or receivers are allowed to cooperate.
Specifically, we focus on the stability of the grand coalition and seek to
understand if the GC also maximizes the utilities of all the users. For
specific encoding and decoding schemes, we model the maximum achievable
information-theoretic rate as a measure of a user's utility. The encoding and
decoding schemes also determine the manner in which the rate gains can be
apportioned between the cooperating users in a coalition. Coalitional games
are classified into two types based on the apportioning of gains among users
in a coalition \cite[Section IV]{Rubenstein:acigt}: i) a \textit{transferable}
\textit{utility }(TU)\textit{\ }game where the total rate achieved is
apportioned arbitrarily between the users in a coalition subject to
feasibility constraints and ii) a \textit{non-transferable} \textit{utility}
(NTU) game where the apportioning strategies have additional constraints that
prevent arbitrary apportioning.%

In \cite{MSM:01,MSM:02}, we apply results from information theory and
TU\ games to study the stable coalition structure when only receivers in an IC
cooperate by jointly decoding their received signals. We show that the GC of
receivers is the stable sum-rate maximizing coalition structure. On the other
hand, for the case where the receivers cooperate using linear multiuser
detectors, we show that the GC is always the stable coalition for the MMSE
detector and is stable only in the high signal-to-noise ratio (SNR) regime for
the decorrelating detector. We briefly review our results in Section
\ref{Sec_4}.

In this paper, we study the formation of stable coalitions when transmitters
are allowed to cooperate in a $K$-link IC. The cooperative strategies and rate
regions for a $2$-link IC with varying degrees of transmitter and receiver
cooperation is studied in \cite{HostMadsen:01,NgJindalMitraGS:01} and the
references therein. For a $K$-link IC, there is a combinatorial explosion in
the ways in which the transmitters can cooperate. Thus, knowledge of the
stable coalition structures can be useful in choosing the appropriate
cooperative strategies. We assume that the $K$ receivers jointly decode their
received signals thus simplifying the IC to a multi-access (MAC) channel with
a multi-antenna receiver. We also assume that transmitters in a coalition have
no knowledge of the transmission strategies of the users outside. We model the
lack of transmit information between competing coalitions as a
\textit{jamming} game, i.e., we assume that each coalition determines its
stability by assuming worst case jamming interference from other coalitions.
We first study the TU\ game that results when the transmitters in a coalition
cooperate perfectly, i.e., each transmitter has perfect knowledge of the
messages of the other transmitters in its coalition. We prove that the game is
\textit{cohesive}~\cite[chap. 13]{Rubenstein:acigt}, i.e., the largest
$K$-user sum-rate is achieved by the GC. This allows us to show that the GC is
the only viable stable coalition structure \cite[p. 258]{Rubenstein:acigt},
i.e., no stable coalition structure exists when the GC is not stable. Finally,
using examples we demonstrate that the GC is not always stable and that the
stability depends on the relative strengths of the user channels to the destination.

We also study the NTU game that results when all the transmitters in a
coalition decode and jointly forward a part of their message streams via a
\textit{partial decode-and-forward} (PDF) strategy
\cite{cap_theorems:FMJWthesis, LSNGKRNBM:01}. We assume perfectly cooperating
co-located receivers with fixed channel gains thus simplifying the IC to a
cooperative MAC. Motivated by the results for the perfect transmitter
cooperation game, we focus on a class of channels where all the users are
\textit{clustered}, i.e., their inter-user links are stronger than the links
between the users and the destination. For this class, we prove that the
achievable rate region is maximized when transmitters in a coalition decode
all messages from one another thus generalizing the results for a two-user
cooperative MAC in \cite[Proposition 1]{UlukusKaya:01}. However, using
examples, we show that when the jamming is weak, users may have incentives to break away from the cluster, i.e.,  the game may not be
cohesive. These results for clustered users also point to the fact that for
the general class of channels with arbitrary inter-user links the game may not
be cohesive in general.

This paper is organized as follows. In Section \ref{Sec_2} we provide an
overview of coalitional game theory. In section \ref{Sec_3} we introduce the
system models. In Section \ref{Sec_4} we review our results on receiver
cooperation. In Section \ref{Sec_5}, we study transmitter cooperation as a
coalitional game using two different cooperation models. We conclude in
Section \ref{Sec_6}.

\section{\label{Sec_2}Coalitional Game Theory for Receiver and Transmitter
Cooperation}

We use the framework of coalitional game theory to determine the stable rate
maximizing cooperative coalitions in a wireless network. To determine
stability one must in general take into account the fact that the rate
achieved by a coalition is also affected by the actions of the users outside
the coalition. However, determining the stable coalition structures for such a
general model is not straightforward \cite[p. 258]{Rubenstein:acigt}. Thus, it
is common practice to assume that a game is in \textit{characteristic function
form} (CFF), i.e., the utilities achieved by the users in a coalition are
unaffected by those outside it~\cite{Aumann:01}.

When only receivers cooperate, the game is in CFF. This is due to the fact
that the transmitters in these models do not cooperate. In fact, for a fixed
encoding at the transmitters, the rate achieved by any coalition only depends
on the combined interference presented by the users outside the coalition and
not on the coalition structures to which they belong. On the other hand, the
games resulting from both kinds of transmitter cooperation models are not in
CFF because the cooperative strategies of users outside a coalition affects
the rates achieved by the members of a coalition. We convert the game to a CFF
by considering a \textit{jamming game}, i.e., we assume that a coalition
assumes that the users outside cooperate to act as worst case jammers.

Games in CFF can be further categorized as TU and NTU games depending on
whether the cooperative gains are divided arbitrarily or in a constrained
manner, respectively. We define both games and their properties below.

\begin{definition}
A coalitional game with transferable utility $\langle\mathcal{K},v\rangle$ is
defined as \cite[Chap. 13]{Rubenstein:acigt}

\begin{itemize}
\item a finite set of users $\mathcal{K}$ ,

\item a value $v(\mathcal{S})\in\mathbb{R}_{+}$ for all $\mathcal{S}%
\subseteq\mathcal{K}$ with $v\left(  \{\phi\}\right)  =0$.
\end{itemize}
\end{definition}

A coalition structure is a partition of the set $\mathcal{K}$, and thus the
number of coalition structures, i.e., the number of possible partitions of
$\mathcal{K}$, grows exponentially with $K$ \cite{SLAST:01}. In fact, it has
been shown that finding the sum-rate maximizing coalition structure is an
$NP$-complete problem~\cite{SLAST:01}. To this end, the following properties
of a TU game greatly simplify such a search.

\begin{definition}
\label{cohesive} A coalitional game with transferable utility is said to be
\textit{cohesive} if the value of the grand coalition formed by the set of all
users $\mathcal{K}$ is at least as large as the sum of the values of any
partition of $\mathcal{K}$, i.e.
\begin{equation}
\sum_{n=1}^{N}v(\mathcal{S}_{n})\leq v(\mathcal{K}) \label{Coh_defn}%
\end{equation}
for any partition $(\mathcal{S}_{1},\ldots,\mathcal{S}_{N})$ of $\mathcal{K}$
where $2\leq N\leq K$.
\end{definition}

\begin{remark}
\label{Cohesive_Rem}A TU game that is cohesive has the GC as the optimal
coalition structure \cite[p. 258]{Rubenstein:acigt}, i.e., the sum of the
utilities of all the users is maximum. This follows from the fact that all
other coalition structures will be unstable as every user has an incentive to
join the GC and benefit from a redistribution of total utility.
\end{remark}

In addition to being cohesive, a TU coalitional game can also be superadditive
which is defined as follows.

\begin{definition}
\label{SupAdd}A coalitional game with transferable payoff is said to be
superadditive if for any two disjoint coalitions $\mathcal{S}_{1}$ and
$\mathcal{S}_{2}$, we have
\begin{equation}
v(\mathcal{S}_{1}\cup\mathcal{S}_{2})\geq v(\mathcal{S}_{1})+v(\mathcal{S}%
_{2}).\newline\label{SupAdd_defn}%
\end{equation}

\end{definition}

\begin{remark}
Comparing (\ref{Coh_defn}) and (\ref{SupAdd_defn}), we see that
superadditivity requires the cohesive property to hold for any two disjoint
subsets of $\mathcal{K}$ with respect to their union.
\end{remark}

We refer to a vector describing the share of the rate (payoffs) received by
the members (players) of a coalition as a \emph{payoff vector}.

\begin{definition}
\label{Def_S_feasible}For any coalition $\mathcal{S}$, a vector \underline
{$x$}$_{\mathcal{S}}=(x_{m})_{m\in\mathcal{S}}$ of real numbers is a
$\mathcal{S}$-\textit{feasible payoff vector} if $x(\mathcal{S})=\sum
_{m\in\mathcal{S}}x_{m}=v(\mathcal{S})$. The $\mathcal{K}$-feasible payoff
vector is referred to as a \textit{feasible payoff profile}.
\end{definition}

Of all possible coalition structures, the ones that are stable are of most
interest. Further, due to the complexity of finding stable coalition
structures for non-cohesive games where the GC does not achieve the largest
value, coalitional games that are cohesive are the easiest to study. For
wireless networks, such games also optimize the spectrum utilization. In the
following definition, we assume that the game is cohesive and thus the GC is
the only possible stable coalition.

\begin{definition}
\label{Def_core}The \textit{core}, $C(v)$, of a coalitional game with
transferable payoff, $\langle\mathcal{K},v\rangle$, is the set of feasible
payoff profiles \underline{$x$}$_{\mathcal{K}}$ for which there is no
coalition $\mathcal{S}\subset\mathcal{K}$ and a corresponding $\mathcal{S}%
$-feasible payoff vector \underline{$y$}$_{\mathcal{S}}=(y_{m})_{m\in
\mathcal{S}}$ such that $y_{m}>x_{m}$ for all $m\in\mathcal{S}$.
\end{definition}

For TU games, Definition \ref{Def_core} simplifies to the condition that the feasible payoff
profiles \underline{$x$}$_{\mathcal{K}}$ in the core satisfy%
\begin{align}
&
\begin{array}
[c]{cc}%
x(\mathcal{S})=\sum_{m\in\mathcal{S}}x_{m}\geq v(\mathcal{S}) & \text{for all
}\mathcal{S}\subset\mathcal{K}%
\end{array}
\label{Core_consts1}\\
&
\begin{array}
[c]{cc}%
x(\mathcal{K})=\sum_{m\in\mathcal{K}}x_{m}=v(\mathcal{K}). &
\end{array}
\label{Core_consts2}%
\end{align}
This follows from the fact that in a game with transferable payoff if there
exists a coalition $\mathcal{S}$ with $v(\mathcal{S})>x(\mathcal{S})$ then we
can always find a $\mathcal{S}$-feasible payoff vector \underline{$y$%
}$_{\mathcal{S}}$ such that $y_{k}>x_{k}$, for all $k\in\mathcal{S}$. Such an
assignment can result, for instance, when the $\mathcal{S}$-feasible payoff
vector \underline{$y$}$_{\mathcal{S}}$ is constructed by assigning to each
link $k\in\mathcal{S}$, the payoff $x_{k}$ and then uniformly apportioning the
surplus payoff $v(\mathcal{S})-x(\mathcal{S})$ between links in $\mathcal{S}$.
\ We use this equivalent definition to determine the stability of the core.
Finally, we remark that determining the non-emptiness of the core simplifies
to determining whether the linear program defined by the inequalities in
(\ref{Core_consts1}) and (\ref{Core_consts2}) is feasible.

We formally define an NTU game and its properties below \cite[p.
268]{Rubenstein:acigt}.

\begin{definition}
\label{NTU_def}A coalitional game with non-transferable utility $\langle
\mathcal{K},\mathcal{V}\rangle$ consists of

\begin{itemize}
\item A finite set $\mathcal{K}$ of $K$ players,

\item A set function $\mathcal{V}:\mathcal{S\rightarrow}\mathbb{R}_{+}^{K}$
such that for all $\mathcal{S}\subseteq\mathcal{K}$

\begin{itemize}
\item $\mathcal{V}(\phi)=\phi$ (normalized)

\item $\mathcal{V}(\mathcal{S})$ is a non-empty closed subset of
$\mathbb{R}_{+}^{K}$ such that the components of the rate tuples in
$\mathcal{V}(\mathcal{S})$ whose indices correspond to players not in
$\mathcal{S}$ can be arbitrary,

\item for any length-$K$ vectors $\underline{x}\in\mathcal{V}(\mathcal{K})$
and $\underline{y}\in\mathbb{R}_{+}^{K}$ with entries $y_{k}\leq x_{k}$, for
all $k$, we have $\underline{y}\in\mathcal{V}(\mathcal{K})$ (comprehensive).
\end{itemize}
\end{itemize}
\end{definition}

\begin{definition}
\label{NTU_Coh}An NTU coalitional game $\langle\mathcal{K},\mathcal{V}\rangle$
is cohesive if and only if
\begin{equation}
\bigcap_{n=1}^{N}\mathcal{V}(\mathcal{S}_{n})\subseteq\mathcal{V}(\mathcal{K})
\label{NTU_Cohesive}%
\end{equation}
where $\{\mathcal{S}_{1},\mathcal{S}_{2},\ldots,\mathcal{S}_{N}\}$ is any
partition of $\mathcal{K}$ where $2\leq N\leq K$.
\end{definition}

As with TU\ games, we focus on the stability of the GC and define a core of a
NTU\ game that is cohesive.

\begin{definition}
The core $\mathcal{C}(\mathcal{K},\mathcal{V})$ of an NTU coalitional game
$\langle\mathcal{K},\mathcal{V}\rangle$ is the set of payoff vectors
$\underline{x}\in\mathcal{V}(\mathcal{K})$ such that there is no coalition
$\mathcal{S}$ and payoff vector $\underline{y}\in\mathcal{V}(\mathcal{S})$
such that $y_{k}>x_{k}$ for all $k\in\mathcal{S}$.
\end{definition}

\section{\label{Sec_3}Channel and Cooperation Models}

\subsection{\label{SubSec2_1}Channel\ Model}%

\begin{figure}
[ptb]
\begin{center}
\includegraphics[
height=2.1707in,
width=3.1324in
]%
{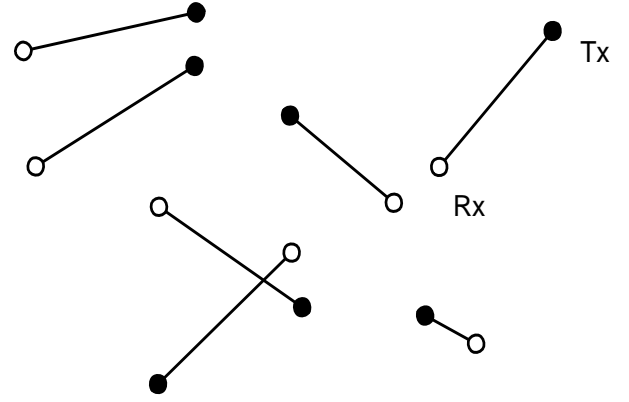}%
\caption{An interference channel with $K$ transmit-receive links.}%
\label{Fig_IC}%
\end{center}
\end{figure}
Our network consists of $K$ transmitter-receiver pairs (links), indexed by the
set $\mathcal{K}=\{1,\ldots,K\}$ \cite{ABC:IntCh} (see Fig. \ref{Fig_IC}). We
model each link as an additive white Gaussian noise channel with fixed channel
gains. The received signal at receiver $m$ is given by
\begin{equation}%
\begin{array}
[c]{cc}%
Y_{m}=\sum_{k=1}^{K}\sqrt{h_{m,k}}X_{k}+Z_{m} & m\in\mathcal{K}%
\end{array}
\label{first_equation_IC}%
\end{equation}
where $h_{m,k}^{1/2}$ is the channel gain between transmitter $k$ and receiver
$m$. The noise entries $Z_{m}\sim\mathcal{CN}(0,1)$, for all $m$, are
independent, identically distributed (i.i.d), proper complex zero-mean
unit-variance Gaussian random variables. The transmit power at transmitter $k$
is constrained as
\begin{equation}%
\begin{array}
[c]{cc}%
E|X_{k}|^{2}\leq P_{k} & \text{for all }k\in\mathcal{K}\text{.}%
\end{array}
\label{Input_power_constraint}%
\end{equation}
We assume that the transmitters employ Gaussian signaling subject to
(\ref{Input_power_constraint}). For the case where the receivers are
co-located, our model simplifies to a MAC where all the transmitters
communicate with the same destination, denoted as $d$ such that $Y_{d}=Y_{k}$
for all $k$. Finally, we write $X_{\mathcal{S}}=\{X_{k}:k\in\mathcal{S}\}$ for
all $\mathcal{S}\subseteq\mathcal{K}$ and $\mathcal{S}^{c}$ as the complement
of $\mathcal{S}$ in $\mathcal{K}$. Finally, throughout the paper, we use the
words user and transmitter interchangeably.

\subsection{\label{Sec_3_2}Cooperation Models}

\paragraph{Receiver cooperation via Joint decoding}

We assume that the receivers that cooperate communicate via noise-free links
and that the transmitters do not cooperate. We assume that a coalition of
cooperating receivers treats signals from transmitters outside the coalition
as interference. For the channel in (\ref{first_equation_IC}), each
non-singleton coalition can thus be modeled as a single-input, multiple-output
Gaussian multiple access channel (SIMO-MAC) whose capacity region is known
\cite{Telatar:01} and achieved by the Gaussian input signaling chosen.

\paragraph{Receiver cooperation using Linear multiuser detectors}

We assume an IC with co-located receivers thereby simplifying the channel to a
single-antenna MAC. We consider a BPSK modulated, synchronized CDMA system
with no power control such that the correlation between any two user signature
sequences is $\rho$. We write the signal at the receiver as \cite[p.
19]{Verdu:MUD}%
\begin{equation}%
\begin{array}
[c]{cc}%
y(t)=\sum\limits_{k=1}^{K}\sqrt{P}h_{k}b_{k}s_{k}\left(  t\right)  +\sigma
n\left(  t\right)  , & t\in\left[  0,T\right]
\end{array}
\end{equation}
where $P$ is the common transmit power of all users, $h_{k}$ is the channel
gain from user $k$ to the receiver, $b_{k}\in\left\{  +1,-1\right\}  $ is the
bit transmitted by user $k$ in the bit interval $[0,T]$, $s(t)$ is the
signature sequence of user $k$, and $n(t)$ is an additive white Gaussian noise
process with unit variance. The received signal is filtered through a bank of
$K$ matched filters to obtain a $K\times1$ received signal vector
\cite{Verdu:MUD}
\begin{equation}
\mathbf{y}=\mathbf{RAb}+\mathbf{n} \label{Rx_MUD}%
\end{equation}
where $\mathbf{R\in}\mathcal{R}^{K\times K}$ is a signature sequence cross
correlation matrix, $\mathbf{A}$ is a diagonal matrix containing the received
amplitudes $\sqrt{P}h_{k}$, for all $k$, $\mathbf{b}$ is an $K\times1$ vector
of transmitted bits, and $\mathbf{n}$ is a Gaussian random vector with zero
mean and covariance $\sigma^{2}\mathbf{R}$.

\textit{Transmitter Cooperation}: We study two models for transmitter
cooperation in a $K$-link IC. In both cases, we assume that the receivers of
all the links jointly decode (see Fig. \ref{Fig_TxCoop}). \ Further, for
simplicity, under PDF, we assume co-located receivers thus simplifying the IC
to a cooperative MAC. Finally, in both cases, we assume that each coalition is
affected by worst-case jamming by competing coalitions.\ 

\paragraph{Perfect cooperation}

For perfect transmitter cooperation each non-singleton coalition can be
modeled as a multi-input, $K$-output MIMO channel with per-antenna power constraints. The transmitters in a
coalition maximize their MIMO sum-capacity \cite{Telatar:01} subject to worst
case jamming from other coalitions.

\paragraph{Partial decode-and-forward}

We consider a MAC where a coalition of transmitters cooperate via a PDF scheme
\cite{cap_theorems:FMJWthesis,SEA:01,LSNGKRNBM:01}. We assume full duplex
communications at the cooperating transmitters. The received signals $Y_{d}$
and $Y_{j}$ at the destination and at user $j$, respectively, are
\begin{align}
&
\begin{array}
[c]{cc}%
Y_{d}=\sum_{k=1}^{K}\sqrt{h_{d,k}}X_{k}+Z_{d} &
\end{array}
\label{TxPDF_Yd}\\
&
\begin{array}
[c]{cc}%
Y_{j}=\sum_{k\in\mathcal{K},k\neq j}\sqrt{h_{j,k}}X_{k}+Z_{j} & \text{for
all}\hspace{0.1cm}j\in\mathcal{K}.
\end{array}
\hspace{0.3cm} \label{TxPDF_Yk}%
\end{align}
where $h_{j,k}^{1/2}$ is the channels gain from user $k$ to user $j$, and
$Z_{d}$ and $Z_{j}$ are zero-mean unit variance proper complex Gaussian noise
variables. We focus on a class of \textit{clustered} channels, i.e., a network
where
\begin{equation}%
\begin{array}
[c]{cc}%
h_{m,k}>h_{d,k} & \text{ for all }m\in\mathcal{K},m\not =k.
\end{array}
\label{Cluster_Model}%
\end{equation}
This represents a model where the users are most likely to cooperate to
overcome a relatively poor direct channel to the destination.

\section{\label{Sec_4}Receiver Cooperation}

In \cite{MSM:01, MSM:02}, we determine the stable coalitions when receivers
cooperate in an IC. The cooperation models are described in Section
\ref{Sec_3_2} and we present the results here.

\subsection{Receiver Cooperation via Joint Decoding (TU\ game)}%

\begin{figure}
[ptb]
\begin{center}
\includegraphics[
height=2.8383in,
width=3.2076in
]%
{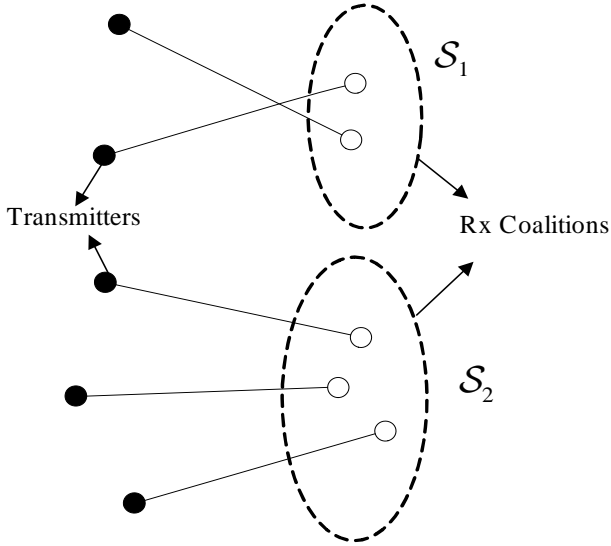}%
\caption{Receiver coalitions formed in a $K$-link IC\ when receivers cooperate
via joint decoding and transmitters do not cooperate.}%
\label{Fig_RxCoop}%
\end{center}
\end{figure}

Consider the TU game that results when cooperating receivers in a $K$-link
IC\ jointly decode their received signals (Fig. \ref{Fig_RxCoop}). For fixed
channel gains, we define the value $v(\mathcal{S})$ of a coalition
$\mathcal{S}$ of links as the maximum information-theoretic sum-rate achieved
by the links in $\mathcal{S}$, i.e., \cite{MSM:01}
\begin{equation}
v(\mathcal{S})=\max\nolimits_{\underline{R}_{\mathcal{S}}\in\mathcal{C}%
_{\mathcal{S}}}{\sum\nolimits_{i\in\mathcal{S}}R_{i}}=\max
\nolimits_{P_{X_{\mathcal{S}}}}I(X_{\mathcal{S}};Y_{\mathcal{S}})
\label{Value_IntCh_Coal_Game}%
\end{equation}
where $\underline{R}_{\mathcal{S}}=(R_{i})_{i\in\mathcal{S}}$ is the vector of
rates for links in $\mathcal{S}$ and $\mathcal{C}_{\mathcal{S}}$ is the
capacity region of the SIMO-MAC formed by the links in $\mathcal{S}$. For
the\ white Gaussian channel considered, the input distribution
$P_{X_{\mathcal{S}}}$ maximizing (\ref{Value_IntCh_Coal_Game}) is zero-mean
independent Gaussian signaling at each transmitter in $\mathcal{S}$ with
variance set to the maximum transmit power in (\ref{Input_power_constraint}).
The value $v(\mathcal{S})$ of a coalition $\mathcal{S}$ can be apportioned
between its members in any arbitrary manner. Depending on its allocated share
of $v(\mathcal{S})$, a receiver may decide to break away from the coalition
$\mathcal{S}$ and join another coalition where it achieves a greater rate. For
this model, we prove the following results (see \cite{MSM:01}).

\begin{theorem}
The grand coalition maximizes spectrum utilization in the joint decoding
receiver cooperation coalitional game.
\end{theorem}

\begin{proof}
From definition \ref{SupAdd} for a superadditive game, the sum-rate of all
links is maximized by the grand coalition. Since maximizing the sum-rate is
equivalent to maximizing the utilization of the shared spectrum, we only need
to show that the value of a coalition for this receiver cooperation
coalitional game is a superadditive function.\newline

Consider two coalitions $\mathcal{S}_{1}$ and $\mathcal{S}_{2}$ such that
$\mathcal{S}_{1}\cap\mathcal{S}_{2}=\phi$. In order to prove that
$v(\mathcal{S})$ is superadditive, we need to show that
\begin{equation}
I(X_{\mathcal{S}_{1}\cup\mathcal{S}_{2}};Y_{\mathcal{S}_{1}\cup\mathcal{S}%
_{2}})\geq I(X_{\mathcal{S}_{1}};Y_{\mathcal{S}_{1}})+I(X_{\mathcal{S}_{2}%
};Y_{\mathcal{S}_{2}}) \label{SupAdd_valfunc}%
\end{equation}
We expand $I(X_{\mathcal{S}_{1}\cup\mathcal{S}_{2}};Y_{\mathcal{S}_{1}%
\cup\mathcal{S}_{2}})$ as%
\begin{multline}
I(X_{\mathcal{S}_{1}\cup\mathcal{S}_{2}};Y_{\mathcal{S}_{1}\cup\mathcal{S}%
_{2}})=I(X_{\mathcal{S}_{1}};Y_{\mathcal{S}_{1}})+I(X_{\mathcal{S}_{1}%
};Y_{\mathcal{S}_{2}}|Y_{\mathcal{S}_{1}})\label{Th1_expand_MI}\\
+I(X_{\mathcal{S}_{2}};Y_{\mathcal{S}_{2}}|X_{\mathcal{S}_{1}}%
)+I(X_{\mathcal{S}_{2}};Y_{\mathcal{S}_{1}}|Y_{\mathcal{S}_{2}},X_{\mathcal{S}%
_{1}})
\end{multline}
Further expanding $I(X_{\mathcal{S}_{2}};Y_{\mathcal{S}_{2}}|X_{\mathcal{S}%
_{1}})$, we have
\begin{align}
I(X_{\mathcal{S}_{2}};Y_{\mathcal{S}_{2}}|X_{\mathcal{S}_{1}})  &
=H(X_{\mathcal{S}_{2}})-H(X_{\mathcal{S}_{2}}|Y_{\mathcal{S}_{2}%
},X_{\mathcal{S}_{1}})\label{Th1_expand_I}\\
&  \geq I(X_{\mathcal{S}_{2}};Y_{\mathcal{S}_{2}}) \label{Th1_expand_I_2}%
\end{align}
where (\ref{Th1_expand_I}) follows from the independence of the transmitter
signals and the inequality in (\ref{Th1_expand_I_2}) from the fact that
conditioning reduces entropy. Finally, comparing (\ref{Th1_expand_MI}) with
(\ref{SupAdd_valfunc}) and using the fact that mutual information is
non-negative, we have that the joint decoding receiver cooperation coalitional
game is superadditive.
\end{proof}

\begin{theorem}
The GC is the stable coalition structure that maximizes the spectrum
utilization in the interference channel with jointly decoding cooperating receivers.
\end{theorem}

\begin{proof}
Since the interference channel coalitional game is superadditive, we need only
consider the definition of the core in the context of the grand coalition. Any
feasible payoff profile $\underline{R}_{\mathcal{K}}=(R_{k})_{k\in\mathcal{K}%
}$ that lies in the capacity region, $\mathcal{C}_{\mathcal{K}}$, of a
SIMO-MAC with $K$ independent transmitters and a $K$-antenna receiver
satisfies the inequalities
\begin{equation}%
\begin{array}
[c]{cc}%
\sum\limits_{k\in\mathcal{S}}R_{k}\leq I(X_{\mathcal{S}};Y_{\mathcal{K}%
}|X_{\mathcal{S}^{c}}) & \forall\mathcal{S}\subseteq\mathcal{K}.
\end{array}
\label{Th2_SIMOMAC_Capregion}%
\end{equation}
For Gaussian MIMO-MAC channels, the bounds in (\ref{Th2_SIMOMAC_Capregion})
are maximized by independent Gaussian signaling at the transmitters. We claim
that every feasible payoff profile $\underline{R}_{\mathcal{K}}$ on the dominant face of the capacity region 
$\mathcal{C}_{\mathcal{K}}$ lies in the core. By the equivalent definition of
the core, in order to prove that a $\underline{R}_{\mathcal{K}}$ satisfying
(\ref{Th2_SIMOMAC_Capregion}) lies in the core, we need to show that
\begin{equation}%
\begin{array}
[c]{cc}%
\sum\limits_{k\in\mathcal{S}}R_{k}\geq v(\mathcal{S}) & \forall~\mathcal{S}%
\subseteq\mathcal{K}%
\end{array}
\end{equation}
Since $\underline{R}_{\mathcal{K}}$ is a feasible payoff profile, i.e., $%
{\textstyle\sum_{k\in\mathcal{K}}}
R_{k}=v(\mathcal{K})$, we have\newline%
\begin{equation}
\sum_{k\in\mathcal{K}}R_{k}=\sum_{k\in\mathcal{S}}R_{k}+\sum_{k\in
\mathcal{S}^{c}}R_{k}=I(X_{\mathcal{K}};Y_{\mathcal{K}}). \label{Th2_RKsum}%
\end{equation}
We rewrite (\ref{Th2_RKsum}) above as
\begin{align}
\sum_{k\in\mathcal{S}}R_{k}  &  =I(X_{\mathcal{K}};Y_{\mathcal{K}})-\sum
_{k\in\mathcal{S}^{c}}R_{k}\\
&  \geq I(X_{\mathcal{S}},X_{{\mathcal{S}}^{c}};Y_{\mathcal{K}}%
)-I(X_{\mathcal{S}^{c}};Y_{\mathcal{K}}|X_{\mathcal{S}}%
)\label{Th2_coreproof_chainrule}\\
&  =I(X_{\mathcal{S}};Y_{\mathcal{S}},Y_{{\mathcal{S}}^{c}}%
)\label{Th2_coreproof_CR_result}\\
&  =I(X_{\mathcal{S}};Y_{\mathcal{S}})+I(X_{\mathcal{S}};Y_{{\mathcal{S}}^{c}%
}|Y_{\mathcal{S}})\\
&  \geq I(X_{\mathcal{S}};Y_{\mathcal{S}}) \label{Th2_coreproof_finalineq}%
\end{align}
where the inequality in (\ref{Th2_coreproof_chainrule}) follows from
(\ref{Th2_SIMOMAC_Capregion}) assuming optimal Gaussian signaling at the
transmitters, (\ref{Th2_coreproof_CR_result}) follows from applying the chain
rule for mutual information in (\ref{Th2_coreproof_chainrule}), and
(\ref{Th2_coreproof_finalineq}) follows from the non-negativity of mutual
information. Thus, we have
\begin{equation}
\sum_{k\in\mathcal{S}}R_{k}\geq I(X_{\mathcal{S}};Y_{\mathcal{S}%
})=v(\mathcal{S})
\end{equation}
The above inequality implies that every point on the \emph{dominant face of
}$\mathcal{C}_{\mathcal{S}}$, i.e., on the plane that maximizes the sum rate
of all transmitters, corresponds to a feasible rate payoff profile that lies
in the core. Thus, the core for the interference channel coalitional game is
not only non-empty but is, in general, also non-unique.
\end{proof}

\subsection{Receiver Cooperation using Multiuser Detectors (NTU\ game)}

\begin{figure}
[ptb]
\begin{center}
\includegraphics[
height=1.4754in,
width=3.4013in
]%
{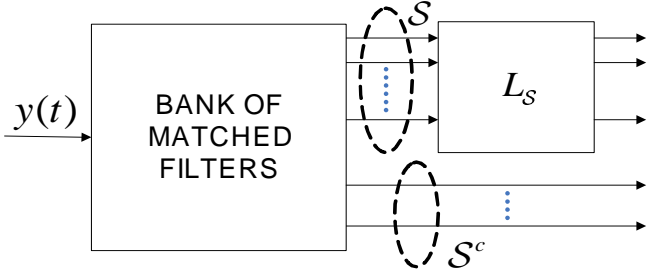}%
\caption{Coalition of links for a decorrelating detector coalitional game.}%
\label{Fig_MUD}%
\end{center}
\end{figure}

In \cite{MSM:02}, we study the stability of the coalitional game that results
when the co-located receivers in an IC use a linear multiuser detector (MUD)
to cooperatively process their matched filter signals \cite[Chaps. 5,
6]{Verdu:MUD}. As described in Section \ref{Sec_3_2}, the transmitters use
random signature sequences to transmit binary signals. We consider a
decorrelating \cite[Chap. 5]{Verdu:MUD} and a MMSE detector \cite[Chap.
6]{Verdu:MUD} and in both cases~determine the SNR regimes for which the GC is
the stable sum-rate maximizing coalition structure. An example of a coalition
of multiuser detectors in shown in Fig. \ref{Fig_MUD}.%

For any coalition $\mathcal{S}\subset\mathcal{K}$, the received signal vector
for this coalition is given by
\begin{equation}
\mathbf{y}_{\mathcal{S}}=\mathbf{R}_{\mathcal{S}}\mathbf{A}_{\mathcal{S}%
}\mathbf{b}_{\mathcal{S}}+\mathbf{R}_{\mathcal{S}^{c}}\mathbf{A}%
_{\mathcal{S}^{c}}\mathbf{b}_{\mathcal{S}^{c}}+\mathbf{n}_{\mathcal{S}}%
\end{equation}
where $\mathbf{R}_{\mathcal{S}}$ is the cross correlation matrix of the
transmit signature sequences in $\mathcal{S}\subseteq\mathcal{K}$,
$\mathbf{A}_{\mathcal{S}}$ is a diagonal matrix containing the received
amplitudes $\sqrt{P}h_{k}$ for all $k\in\mathcal{S}$, $\mathbf{b}%
_{\mathcal{S}}$ is the vector of bits from transmitters in $\mathcal{S}$, and
$\mathbf{n}_{\mathcal{S}}$ is a random Gaussian vector with zero mean and
covariance matrix $\sigma^{2}\mathbf{R}_{\mathcal{S}}$. The $\left\vert
\mathcal{S}\right\vert \times\left\vert \mathcal{S}^{c}\right\vert $ matrix
$\mathbf{R}_{\mathcal{S}^{c}}$ contains the cross correlations between the
signature sequences of users in $\mathcal{S}$ and $\mathcal{S}^{c}$, i.e.,
$\left(  \mathbf{R}_{\mathcal{S}^{c}}\right)  _{ij}=\rho$, for all
$i=1,2,\ldots,\left\vert \mathcal{S}\right\vert $ and $j=1,2,\ldots
,K-\left\vert \mathcal{S}\right\vert $. The $\left\vert \mathcal{S}%
^{c}\right\vert \times\left\vert \mathcal{S}^{c}\right\vert $ diagonal matrix
$\mathbf{A}_{\mathcal{S}^{c}}$ and the $\left\vert \mathcal{S}^{c}\right\vert
$-length vector $\mathbf{b}_{\mathcal{S}^{c}}$ contains the amplitudes and
bits, respectively, of transmitters in $\mathcal{S}^{c}$.

A multiuser detector for the coalition $\mathcal{S}$ applies a linear
transformation $\mathbf{L}_{\mathcal{S}}$ and the resulting vector
$\mathbf{L_{\mathcal{S}}y}_{\mathcal{S}}$ is used to decode the bits from the
transmitters in $\mathcal{S}$. For the decorrelating receiver, $\mathbf{L}%
_{\mathcal{S}}=\mathbf{R}_{\mathcal{S}}^{-1}$ and for the MMSE receiver,
$\mathbf{L_{\mathcal{S}}=}\left(  \mathbf{R_{\mathcal{S}}+\sigma}%
^{2}\mathbf{A}_{\mathcal{S}}^{-2}\right)  ^{-1}$. Links within a coalition
benefit from interference suppression offered by their MUD. The coalitional
games for both detectors are NTU games since linear MUDs achieve a specific
rate tuple for each user in the coalition. Finally, for both detectors we
assume that the rate achieved by each link is a monotonically increasing
function of its signal-to-interference noise ratio (SINR) at the receiver.

\begin{theorem}
[\cite{MSM:02}]The grand coalition is always the stable and sum-rate
maximizing coalition for the receiver cooperation game using a MMSE detector.
\end{theorem}

\begin{proof}
For a coalition $\mathcal{S}$, the linear MMSE\ receiver minimizes both the
noise and the interference for the links in $\mathcal{S}$ by applying the
linear transformation $L_{\mathcal{S}}=\left[  \mathbf{R}_{\mathcal{S}}%
+\sigma^{2}\mathbf{A}_{\mathcal{S}}^{2}\right]  ^{-1}.$ It can be shown that
the SINR$\ \gamma_{k}\left(  \mathcal{S}\right)  $ of transmitter $k$
belonging to the coalition $\mathcal{S}$, for all $k\in\mathcal{S}$, is
\cite{YLAE:01}%
\begin{equation}
\gamma_{k}\left(  \mathcal{S}\right)  =\frac{\left[ \left(  \mathbf{L}%
_{\mathcal{S}}\mathbf{R}_{\mathcal{S}}\right)  _{kk}\right]^{2}h_{k}%
^{2}P}{\left(
\begin{array}
[c]{c}%
\sigma^{2}\left(  \mathbf{L}_{\mathcal{S}}\mathbf{R}_{\mathcal{S}}%
\mathbf{L}_{\mathcal{S}}\right)  _{kk}+\rho^{2}\left[  \left(  \mathbf{L}%
_{\mathcal{S}}\mathbf{e}_{\mathcal{S}}\right)  _{k}\right]  ^{2}%
\sum_{j\not \in \mathcal{S}}h_{j}^{2}P\\
+\sum_{j\in\mathcal{S},j\not =k}\left[  \left(  \mathbf{L}_{\mathcal{S}%
}\mathbf{R}_{\mathcal{S}}\right)  _{kj}\right]  ^{2}h_{j}^{2}P
\end{array}
\right)  } \label{RxMUD_SINR}%
\end{equation}
where $\mathbf{e}_{\mathcal{S}}$ is a vector of length $\left\vert
\mathcal{S}\right\vert $ with entries $e_{k}=1$ for all $k$. The second and
third terms in the denominator of (\ref{RxMUD_SINR}) are the interference
presented to link $k$ from other links outside and within $\mathcal{S}$,
respectively. From (\ref{RxMUD_SINR}) the SINR, and hence, the rate achieved
by every transmitter is maximized when all users are a part of the grand
coalition. Thus, every transmitter would prefer to belong to the grand
coalition where it is not subject to additional interference from
non-cooperating transmitters, i.e., the grand coalition is both sum-rate
maximizing and stable.
\end{proof}

\begin{theorem}
The grand coalition is the stable and sum-rate maximizing coalition in the
high SNR regime for the receiver cooperation game using a decorrelating detector.

\begin{proof}
The SINR $\eta_{k}\left(  \mathcal{S}\right)  $ achieved at the decorrelating
receiver by every transmitter $k$ in the coalition $\mathcal{S}$ is
\cite{YLAE:01}%
\begin{equation}
\eta_{k}\left(  \mathcal{S}\right)  =\frac{h_{k}^{2}P}{\left[  \frac
{\sigma^{2}}{1-\rho}\cdot\frac{1+\rho\left(  \left\vert \mathcal{S}\right\vert
-2\right)  }{1+\rho\left(  \left\vert \mathcal{S}\right\vert -1\right)
}+\left[  \frac{\rho}{1+\rho\left(  \left\vert \mathcal{S}\right\vert
-1\right)  }\right]  ^{2}\sum_{j\not \in \mathcal{S}}h_{j}^{2}P\right]  }
\label{SINR_Decorr}%
\end{equation}
where the first and second terms in the denominator of (\ref{SINR_Decorr}) are
the interference due to other links within and outside the coalition
$\mathcal{S}$, respectively. Recall that the core of a NTU game is the set of
all payoff profiles for which there is no coalition $\mathcal{S}%
\subset\mathcal{K}$ that can achieve a payoff vector $\underline
{R}_{\mathcal{S}}=(R_{k})_{k\in\mathcal{S}}$ such that $R_{k}\left(
\mathcal{S}\right)  >R_{k}\left(  \mathcal{K}\right)  $ for all $k\in
\mathcal{S}$. From (\ref{SINR_Decorr}), we see that the payoff of any link $k$
when it is a part of the grand coalition is
\begin{equation}
\eta_{k}\left(  \mathcal{K}\right)  =\frac{h_{k}^{2}P}{\frac{\sigma^{2}%
}{1-\rho}\cdot\frac{1+\rho\left(  K-2\right)  }{1+\rho\left(  K-1\right)  }}.
\label{SINR_Dec_GC}%
\end{equation}
Further, comparing (\ref{SINR_Decorr}) and (\ref{SINR_Dec_GC}), in the high
SNR regime we have%
\begin{equation}
\lim_{\sigma\rightarrow0}\eta_{k}\left(  \mathcal{S}\right)  <\lim
_{\sigma\rightarrow0}\eta_{k}\left(  \mathcal{K}\right)  .
\end{equation}
Thus, in the high SNR\ regime, the grand coalition is stable as every link
achieves its largest SINR, and hence, rate, when it is a part of the grand
coalition and therefore has no incentive to defect.
\end{proof}
\end{theorem}

\section{\label{Sec_5}Transmitter Cooperation}

\subsection{\label{TxCoop_1}Transmitter Cooperation: Perfect Transmit
Side-Information}%

\begin{figure}
[ptb]
\begin{center}
\includegraphics[
height=3.0735in,
width=3.2975in
]%
{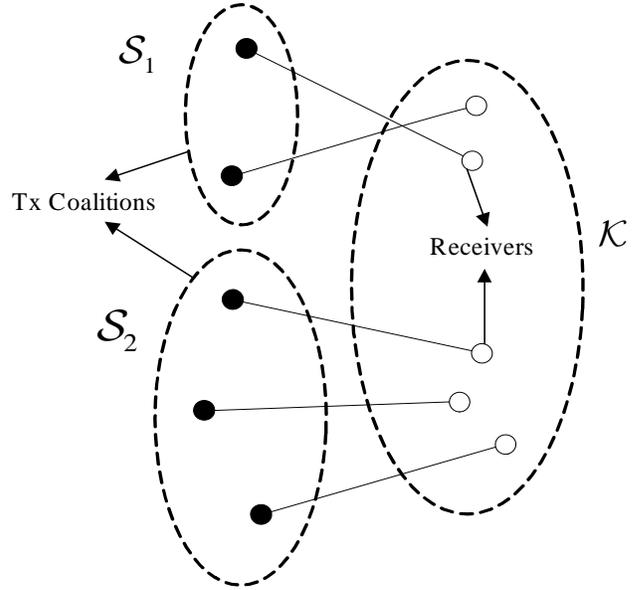}%
\caption{Transmitter coalitions in a $K$-link IC when transmitters cooperate
via noise-free links and all $K$ receivers cooperate. }%
\label{Fig_TxCoop}%
\end{center}
\end{figure}

The $K$ receivers jointly decode their received signals, and thus, can be
considered as a distributed $K$-antenna receiver. For any coalition structure
$(\mathcal{S}_{1},\mathcal{S}_{2},\ldots,\mathcal{S}_{N})$ where $2\leq N\leq
K$, the IC simplifies to a MIMO-MAC with per-antenna power constraints such
that the transmitters in a coalition act as a single transmitter with multiple
antennas (see Fig. \ref{Fig_TxCoop} for $N=2$)$.$ For the GC $(N=1)$ the
cooperative channel further simplifies to a MIMO point-to-point channel with
per antenna power constraints. From (\ref{first_equation_IC}), we write the
$K\times1$ vector of received signals at the $K$ receivers, $\underline
{Y}_{\mathcal{K}}$, as
\begin{equation}
\underline{Y}_{\mathcal{K}}=\sum\nolimits_{n=1}^{N}\mathbf{H}_{\mathcal{S}%
_{n}}\underline{X}_{\mathcal{S}_{n}}+\underline{Z}_{\mathcal{K}}
\label{TxperfCoop_YK}%
\end{equation}
where $\mathbf{H}_{\mathcal{S}_{n}}$ is a $K\times|\mathcal{S}_{n}|$ channel
gains matrix, $\underline{X}_{\mathcal{S}_{n}}$ is an input vector whose
$i^{th}$ entry is the signal transmitted by the $i^{th}$ transmitter in the
coalition $\mathcal{S}_{n}$, and $\underline{Z}_{\mathcal{K}}$ is the noise
vector whose $k^{th}$ entry $Z_{k}$ is the noise at the $k^{th}$ receiver. For
the received signals in (\ref{TxperfCoop_YK}), we obtain the sum-rate achieved
by the coalition $\mathcal{S}_{n}$ as the capacity of a $\left\vert
\mathcal{S}_{n}\right\vert \times K$ MIMO\ channel \cite{Telatar:01} subject
to worst case interference from the users not in $\mathcal{S}_{n}$. This is a
mutual information game \cite[Chap. 10, p. 263]{cap_theorems:CTbook} and thus
the sum-rate of a coalition is both maximized and minimized by Gaussian
signaling at the users in $\mathcal{S}_{n}$ and $\mathcal{S}_{n}^{c}$,
respectively, for all $n$. Further, the rate achieved by transmitters in a
coalition can be arbitrarily apportioned between its users and thus the
transmitter cooperation game is a TU game. We henceforth refer to this game as
a \textit{transmitter cooperation jamming game}.

We write $\mathbf{Q}_{\mathcal{A}}=E[X_{\mathcal{A}}X_{\mathcal{A}}^{\dag}]$
to denote the covariance matrix of the users in $\mathcal{A}$ for all
$\mathcal{A}\subseteq\mathcal{K}$ where $\dag$ denotes the conjugate transpose
of a matrix and $I_{K}$ for the identity matrix of size $K$. For Gaussian
signaling, the value $v(\mathcal{S})$ of a coalition $\mathcal{S}$ of
transmitters is given as%
\begin{align}
v(\mathcal{S})  &  =\min_{\mathbf{Q}_{\mathcal{S}^{c}}}\max_{\mathbf{Q}%
_{\mathcal{S}}}I(X_{\mathcal{S}};Y_{\mathcal{K}})\\
&  =\min\limits_{\mathbf{Q}_{\mathcal{S}^{c}}}\max\limits_{\mathbf{Q}%
_{\mathcal{S}}}{\left\{  \log{\ }\left(  \frac{\left\vert \mathbf{I}%
_{K}{+\mathbf{H}_{\mathcal{K}}\mathbf{Q}_{\mathcal{K}}\mathbf{H}_{\mathcal{K}%
}^{\dag}}\right\vert }{\left\vert {\mathbf{I}}_{K}{+\mathbf{H}_{\mathcal{S}%
^{c}}\mathbf{Q}_{\mathcal{S}^{c}}\mathbf{H}_{\mathcal{S}^{c}}^{\dag}%
}\right\vert }\right)  \right\}  } \label{jammingversion}%
\end{align}
such that the diagonal entries of $\mathbf{Q}_{\mathcal{A}}$ for all
$\mathcal{A}$ are constrained by (\ref{Input_power_constraint}) as%
\begin{equation}%
\begin{array}
[c]{cc}%
(\mathbf{Q}_{\mathcal{A}})_{kk}\leq P_{k} & \text{for all }k\in\mathcal{A}%
\text{.}%
\end{array}
\end{equation}
We use the following proposition on block diagonal matrix multiplication to
further simplify (\ref{jammingversion}).

\begin{proposition}
\label{Prop_TxCoop}The product $\mathbf{AQA}^{\dag}$ for a block diagonal
matrix $\mathbf{Q}$ and $K\times K$ matrix $\mathbf{A}$ simplifies as%
\begin{equation}
\mathbf{AQA}^{\dag}=\mathbf{A}_{\mathcal{S}}\mathbf{Q_{\mathcal{S}}%
A}_{\mathcal{S}}^{\dag}+\mathbf{A_{\mathcal{S}^{c}}Q_{\mathcal{S}^{c}}%
A}_{\mathcal{S}^{c}}^{\dag} \label{TxCoop_Prop_Eqn}%
\end{equation}
where $\mathbf{Q_{\mathcal{S}}}$ and $\mathbf{Q_{\mathcal{S}^{c}}}$ are square
matrices and $\mathbf{A_{\mathcal{S}}}$ and $\mathbf{A_{\mathcal{S}^{c}}}$ are
$K\times\left\vert \mathcal{S}\right\vert $ and $K\times\left\vert
\mathcal{S}^{c}\right\vert $ matrices, respectively, such that%
\begin{equation}%
\begin{array}
[c]{ccc}%
\mathbf{Q}=\left(
\begin{array}
[c]{cc}%
\mathbf{Q_{\mathcal{S}}} & 0\\
0 & \mathbf{Q_{\mathcal{S}^{c}}}%
\end{array}
\right)  & \text{and} & \mathbf{A=}\left(
\begin{array}
[c]{cc}%
\mathbf{A}_{\mathcal{S}} & \mathbf{A}_{\mathcal{S}^{c}}%
\end{array}
\right)  .
\end{array}
\label{TxCoop_Q_bd}%
\end{equation}

\end{proposition}

\begin{proof}
The proof follows simply from expanding $\mathbf{Q}$ and $\mathbf{A}$ as in
(\ref{TxCoop_Q_bd}), respectively, such that
\begin{align}
\mathbf{AQA}^{\dag}  &  =\left(
\begin{array}
[c]{cc}%
\mathbf{A}_{\mathcal{S}} & \mathbf{A}_{\mathcal{S}^{c}}%
\end{array}
\right)  \left(
\begin{array}
[c]{cc}%
\mathbf{Q_{\mathcal{S}}} & 0\\
0 & \mathbf{Q_{\mathcal{S}^{c}}}%
\end{array}
\right)  \left(
\begin{array}
[c]{c}%
\mathbf{A}_{\mathcal{S}}^{\dag}\\
\mathbf{A}_{\mathcal{S}^{C}}^{\dag}%
\end{array}
\right) \\
&  =\left(
\begin{array}
[c]{cc}%
\mathbf{A}_{\mathcal{S}}\mathbf{Q_{\mathcal{S}}} & \mathbf{A}_{\mathcal{S}%
^{c}}\mathbf{Q_{\mathcal{S}^{c}}}%
\end{array}
\right)  \left(
\begin{array}
[c]{c}%
\mathbf{A}_{\mathcal{S}}^{\dag}\\
\mathbf{A}_{\mathcal{S}^{C}}^{\dag}%
\end{array}
\right)
\end{align}
which simplifies to (\ref{TxCoop_Prop_Eqn}).
\end{proof}

Since the transmitted signals of users across competing coalitions
$\mathcal{S}$ and $\mathcal{S}^{c}$ are independent, we use Proposition
\ref{Prop_TxCoop} to simplify the $\log$ expression in (\ref{jammingversion})
as%
\begin{equation}
v(\mathcal{S})=\min\limits_{\mathbf{Q}_{\mathcal{S}^{c}}}\max
\limits_{\mathbf{Q}_{\mathcal{S}}}{\log{\ }\left(  \frac{\left\vert
\mathbf{I}_{K}{+\mathbf{H}_{\mathcal{S}}\mathbf{Q}_{\mathcal{S}}%
\mathbf{H}_{\mathcal{S}}^{\dag}+\mathbf{H}_{\mathcal{S}^{c}}\mathbf{Q}%
_{\mathcal{S}^{c}}\mathbf{H}_{\mathcal{S}^{c}}^{\dag}}\right\vert }{\left\vert
{\mathbf{I}_{K}+\mathbf{H}_{\mathcal{S}^{c}}\mathbf{Q}_{\mathcal{S}^{c}%
}\mathbf{H}_{\mathcal{S}^{c}}^{\dag}}\right\vert }\right)  .}
\label{maxminopt}%
\end{equation}

To simplify the optimization in (\ref{maxminopt}), we use the following two
lemmas on functions of symmetric semi-definite matrices where we write
$\mathbb{S}_{+}^{n}$ to denote the set of such matrices.

\begin{lemma}
[\cite{SDTC:01}]\label{Lemma_convex}The function $f:\mathbb{S}_{+}^{n}%
\mapsto\mathbb{R}$ defined as
\begin{equation}
f(K_{z})=\log{\left(  \left\vert K_{x}+K_{z}\right\vert \left/  \left\vert
K_{z}\right\vert \right.  \right)  }%
\end{equation}
is convex in $K_{z}$ given $K_{x}$ is symmetric positive semi-definite. The
convexity is strict if $K_{x}$ is positive definite.
\end{lemma}

\begin{lemma}
[\cite{SDTC:01}]\label{Lemma_concave}The function $g:\mathbb{S}_{+}^{n}%
\mapsto\mathbb{R}$ defined as
\begin{equation}
g(K_{x})=\log{\left(  \left\vert K_{x}+K_{z}\right\vert \left/  \left\vert
K_{z}\right\vert \right.  \right)  }%
\end{equation}
is strictly concave in $K_{x}$ given $K_{z}$ is symmetric positive definite.
\end{lemma}

We use the preceding Lemmas \ref{Lemma_convex} and \ref{Lemma_concave} to
prove the saddle point property of the transmitter cooperation jamming game.
For ease of exposition, we henceforth write $l(\mathbf{Q}_{\mathcal{S}%
},\mathbf{Q}_{\mathcal{S}^{c}})$ to denote the $\log$ expression in
(\ref{maxminopt}).

\begin{lemma}
\label{Lemma_SadPt}The transmitter cooperation jamming game has a saddle point
solution such that
\begin{equation}
l(\mathbf{Q}_{\mathcal{S}},\mathbf{Q}_{\mathcal{S}^{c}}^{\ast})\leq
l(\mathbf{Q}_{\mathcal{S}}^{\ast},\mathbf{Q}_{\mathcal{S}^{c}}^{\ast})\leq
l(\mathbf{Q}_{\mathcal{S}}^{\ast},\mathbf{Q}_{\mathcal{S}^{c}})
\label{Tx_F_sadpt}%
\end{equation}
and%
\begin{equation}
\max\limits_{\mathbf{Q}_{\mathcal{S}}}\min\limits_{\mathbf{Q}_{\mathcal{S}%
^{c}}}l(\mathbf{Q}_{\mathcal{S}},\mathbf{Q}_{\mathcal{S}^{c}})=\min
\limits_{\mathbf{Q}_{\mathcal{S}^{c}}}\max\limits_{\mathbf{Q}_{\mathcal{S}}%
}l(\mathbf{Q}_{\mathcal{S}},\mathbf{Q}_{\mathcal{S}^{c}})
\end{equation}
where $\mathbf{Q}_{\mathcal{S}}^{\ast}$ and $\mathbf{Q}_{\mathcal{S}^{c}%
}^{\ast}$ are covariance matrices that maximize and minimize $l(\mathbf{Q}%
_{\mathcal{S}},\mathbf{Q}_{\mathcal{S}^{c}})$ in (\ref{maxminopt}), respectively.
\end{lemma}

\begin{proof}
The proof follows from the fact that the transmitter cooperation jamming game
is a mutual information game (see \cite[Chap. 10, p. 263]{cap_theorems:CTbook}%
). Further from Lemmas \ref{Lemma_convex} and \ref{Lemma_concave}, the game
has a saddle point at $(\mathbf{Q}_{\mathcal{S}}^{\ast},\mathbf{Q}%
_{\mathcal{S}^{c}}^{\ast})$ satisfying (\ref{Tx_F_sadpt}) such that a
deviation from the optimal matrix for either $\mathcal{S}$ or $\mathcal{S}%
^{c}$ worsens $l(\mathbf{Q}_{\mathcal{S}},\mathbf{Q}_{\mathcal{S}^{c}})$ from
that coalition's standpoint \cite[Chap. 10, p. 263]{cap_theorems:CTbook}.
\end{proof}

\begin{theorem}
\label{Th_TxC_Cohesive}The transmitter cooperation jamming game is cohesive.
\end{theorem}

\begin{proof}
From Definition \ref{cohesive} and Remark \ref{Cohesive_Rem}, the game is
cohesive when%
\begin{equation}
v(\mathcal{K})\geq\sum\nolimits_{i=1}^{N}v(\mathcal{S}_{i})
\label{cohesiveproof}%
\end{equation}
where $\mathcal{S}_{1},\ldots,\mathcal{S}_{N}$ is any partition of
$\mathcal{K}$, and the value $v(\mathcal{S}_{i})$ of coalition $\mathcal{S}%
_{i}$ is obtained from (\ref{maxminopt}) by setting $\mathcal{S}%
=\mathcal{S}_{i}$. The value $v(\mathcal{K})$ of the GC is given by
(\ref{maxminopt}) with $\mathcal{S}=\mathcal{K}$ and $\mathcal{S}%
^{c}=\emptyset$. Consider a coalition structure $\mathcal{S}_{1}%
,\ldots,\mathcal{S}_{N}$, for any $1<N\leq K$. We expand $I(X_{\mathcal{K}%
};Y_{\mathcal{K}})$ as
\begin{align}
I(X_{\mathcal{K}};Y_{\mathcal{K}})  &  =I(X_{\mathcal{S}_{1}},\ldots
,X_{\mathcal{S}_{N}};Y_{\mathcal{K}})\\
&  \geq\sum\nolimits_{i=1}^{N}I(\mathbf{X}_{\mathcal{S}_{i}};\mathbf{Y}%
_{\mathcal{K}}) \label{TxC_VK_ineq}%
\end{align}
where the inequality in (\ref{TxC_VK_ineq}) follows from chain rule of mutual
information \cite[Theorem 2.5.2]{cap_theorems:CTbook} and the fact that
conditioning does not increase entropy. Consider the block diagonal matrix
$\mathbf{Q}_{\mathcal{K}}^{(bd)}$
\begin{equation}
\mathbf{Q}_{\mathcal{K}}^{(bd)}=\left(
\begin{array}
[c]{cccc}%
\mathbf{Q}_{\mathcal{S}_{1}}^{\ast} & 0 & 0 & \ldots\\
0 & \mathbf{Q}_{\mathcal{S}_{2}}^{\ast} & 0 & \ldots\\
0 & 0 & \ddots & \ldots\\
\vdots & \vdots & \ldots & \mathbf{Q}_{\mathcal{S}_{N}}^{\ast}%
\end{array}
\right)  \label{TxC_QKbd}%
\end{equation}
where $\mathbf{Q}_{\mathcal{S}_{i}}^{\ast}$ is the maximizing covariance
matrix for $v(\mathcal{S}_{i})$ for all $i$ and all partitions. From
(\ref{TxC_QKbd}), the covariance matrix $\mathbf{Q}_{\mathcal{S}_{i}^{c}}$ of
the users in $\mathcal{S}_{i}^{c}$ is obtained from $\mathbf{Q}_{\mathcal{K}%
}^{(bd)}$ by deleting the rows and columns corresponding to users in
$\mathcal{S}_{i}$. In the following inequalities we write $\left(
\cdot\right)  _{\mathbf{Q}_{\mathcal{K}}^{\ast}}$ to denote that the
expression $\left(  \cdot\right)  $ is evaluated at $\mathbf{Q}_{\mathcal{K}%
}^{\ast}$. We lower bound $v(\mathcal{K})$ as%
\begin{align}
v(\mathcal{K})  &  =\left[  I(\mathbf{X}_{\mathcal{K}};\mathbf{Y}%
_{\mathcal{K}})\right]  _{\mathbf{Q}_{\mathcal{K}}^{\ast}}\geq\left[
I(\mathbf{X}_{\mathcal{K}};\mathbf{Y}_{\mathcal{K}})\right]  _{\mathbf{Q}%
_{\mathcal{K}}^{(bd)}}\label{TxC_IE3}\\
&  \geq\left[  \sum\nolimits_{i=1}^{N}I(\mathbf{X}_{\mathcal{S}_{i}%
};\mathbf{Y}_{\mathcal{K}})\right]  _{\mathbf{Q}_{\mathcal{K}}^{(bd)}%
}\label{TxC_IE3a}\\
&  =\sum\nolimits_{i=1}^{N}\log{\left\{  \frac{\left\vert \mathbf{I}%
+\mathbf{H}_{\mathcal{S}_{i}}\mathbf{Q}_{\mathcal{S}_{i}}^{\ast}%
\mathbf{H}_{\mathcal{S}_{i}}^{\dag}+\mathbf{H}_{\mathcal{S}_{i}^{c}%
}{\mathbf{Q}}_{\mathcal{S}_{i}^{c}}\mathbf{H}_{\mathcal{S}_{i}^{c}}^{\dag
}\right\vert }{\left\vert \mathbf{I}+\mathbf{H}_{\mathcal{S}_{i}^{c}%
}{\mathbf{Q}}_{\mathcal{S}_{i}^{c}}\mathbf{H}_{\mathcal{S}_{i}^{c}}^{\dag
}\right\vert }\right\}  }\label{TxC_IE4}\\
&  \geq\sum\nolimits_{i=1}^{N}\log{\left\{  \frac{\left\vert \mathbf{I}%
+\mathbf{H}_{\mathcal{S}_{i}}\mathbf{Q}_{\mathcal{S}_{i}}^{\ast}%
\mathbf{H}_{\mathcal{S}_{i}}^{\dag}+\mathbf{H}_{\mathcal{S}_{i}^{c}}%
\mathbf{Q}_{\mathcal{S}_{i}^{c}}^{\ast}\mathbf{H}_{\mathcal{S}_{i}^{c}}^{\dag
}\right\vert }{\left\vert \mathbf{I}+\mathbf{H}_{\mathcal{S}_{i}^{c}%
}\mathbf{Q}_{\mathcal{S}_{i}^{c}}^{\ast}\mathbf{H}_{\mathcal{S}_{i}^{c}}%
^{\dag}\right\vert }\right\}  }\label{TxC_IE5}\\
&  =\sum\nolimits_{i=1}^{N}v(\mathcal{S}_{i}) \label{TxC_IE6}%
\end{align}
where (\ref{TxC_IE3}) follows from Lemmas \ref{Lemma_concave} and
\ref{Lemma_SadPt}, (\ref{TxC_IE3a}) follow from (\ref{TxC_VK_ineq}),
(\ref{TxC_IE4}) follows from Proposition \ref{Prop_TxCoop} and evaluating the
resulting expression at $\mathbf{Q}_{\mathcal{K}}^{(bd)}$, (\ref{TxC_IE5})
follows from Lemma \ref{Lemma_SadPt}, and (\ref{TxC_IE6}) follows from
(\ref{maxminopt}). Note that the $\mathbf{Q}_{\mathcal{S}_{i}^{c}}^{\ast}$ in
(\ref{TxC_IE5}) is the minimizing matrix in (\ref{maxminopt}) for
$\mathcal{S}=\mathcal{S}_{i}$.
\end{proof}

For cohesive games \cite[p. 258]{Rubenstein:acigt}, the grand coalition is the
only possible stable coalition structure. To determine the stability of the GC
for the transmitter cooperation jamming game, i.e., to verify whether the core
of this game is non-empty, we need to show that the GC is guaranteed to have
at least one stable payoff profile. An analytical proof for the core is
intractable since it requires comparing $K$-dimensional rate regions that are
functions of the channel and power parameters. Instead, using the simple
linear programming interpretation described in Section \ref{Sec_2}, we present
a numerical example that illustrates that the core can be empty.

\begin{example}
Consider a $3$-link IC with perfectly cooperating receivers. All the
transmitters have a maximum power constraint of unity and the channel matrix
$\mathbf{H}_{\mathcal{K}}$ with entries $h_{m,k}$ between the $m^{th}$
receiver and $k^{th}$ transmitter is%
\begin{equation}
\mathbf{H}=\left(
\begin{array}
[c]{ccc}%
0.3019 & 0.3772 & 1.8021\times10^{-2}\\
2.6256\times10^{-8} & 3.1413\times10^{-5} & 2.5662\times10^{-5}\\
2.6893\times10^{-6} & 1.9941\times10^{-3} & 0.8502
\end{array}
\right)  . \label{TxPCoop_H_Eg}%
\end{equation}
From (\ref{Core_consts1}) and (\ref{Core_consts2}) in Section \ref{Sec_2}, for
the $\mathbf{H}$ in (\ref{TxPCoop_H_Eg}), the existence of a core with
non-zero rate tuples $(R_{1},R_{2}$,$\ldots,R_{K})$ is equivalent to the
feasibility of the linear program given by $\sum\nolimits_{k\in\mathcal{S}%
}R_{k}\geq v(\mathcal{S})$ for all $\mathcal{S\subseteq K}$ where
$v(\mathcal{S})$ is defined as in (\ref{maxminopt}). Numerical evaluation
reveals that there does not exist a feasible rate vector where all users
achieve rates larger than what they can achieve outside the GC, i.e., the core
is empty. As a result the GC is not stable since a subset of users that can
achieve better rates as a coalition will break away. Note however, that no
other coalition structure is stable either. This is because users breaking
away can be incentivized with larger payoffs by those users who do not wish to
leave the GC. This in turn will result in a different subset of users
attempting to leave the GC for better rates and thus, the game results in an
oscillatory behavior instead of a single convergent stable structure (see also
\cite[p. 259]{Rubenstein:acigt}). Finally, our numerical analyses lead us to
conjecture that the core will be non-empty, i.e., the GC will be stable, when
the channel gains $h_{m,k}$ as well as the powers $P_{k}$ for all $m$ and $k$
are comparable (see \cite[Chap. 4]{cap_theorems:SuhasMathur} for details).
\end{example}

\begin{remark}
The stability of the grand coalition is equivalent to verifying the
feasibility of the linear program given by (\ref{Core_consts1}) and
(\ref{Core_consts2}). Furthermore, (\ref{Core_consts1}) and
(\ref{Core_consts2}) also determine the set of conditions on the channel gains
and transmit powers required to achieve a non-empty core.
\end{remark}

\subsection{Transmitter Cooperation: Partial Decode-and-Forward (PDF)}

We now seek to understand if relaxing the assumption of perfect noiseless
links between the transmitters can still result in the GC as the only
candidate for the core. We thus consider a clustered model introduced in
equation (\ref{Cluster_Model}) where the full-duplex transmitters have noisy
inter-user channels and the receivers are co-located. For this model, we
consider a PDF strategy, introduced in \cite[Chap. 7]{cap_theorems:FMJWthesis}
for a two-user cooperative MAC, and later extended in \cite{LSNGKRNBM:01} for
$K>2$.

Consider a coalition $\mathcal{S\subseteq K}$ of users that cooperate. In the
PDF strategy, user $k\in\mathcal{S}$ transmits the two new messages
$w_{k,1}\in\{1,2,\ldots,2^{nR_{k,1}}\}$ and $w_{k,2}\in\{1,2,\ldots
,2^{nR_{k,2}}\}$ and a cooperative message $w_{0}\in\{1,2,\ldots,2^{nR_{0}}\}$
where $R_{k,1},$ $R_{k,2},$ and $R_{0}$ are the rates in bits per channel use
at which the messages $w_{k,1}$, $w_{k,2}$, and $w_{0}$ are transmitted,
respectively, and $n$ is the number of channel uses \cite{LSNGKRNBM:01}. The
signal $X_{k}$ transmitted by user $k$ is
\begin{equation}%
\begin{array}
[c]{cc}%
X_{k}=X_{k,d}+V_{k,c}+U & \text{for all }k\in\mathcal{S}%
\end{array}
\label{TxPDF_Xk}%
\end{equation}
where $X_{k,d}$, $V_{k,c},$ and $U$ are zero-mean independent Gaussian random
variables that carry the messages $w_{k,1},$ $w_{k,2}$, and $w_{0}$ and have
variances $p_{k,d}$, $p_{k,c}$, and $p_{k,u}$, respectively, such that the
total power $p_{k}$ at user $k$ subject to (\ref{Input_power_constraint}) is
\begin{equation}%
\begin{array}
[c]{cc}%
p_{k}=p_{k,d}+p_{k,c}+p_{k,u}\leq P_{k} & \text{for all }k\in\mathcal{S}.
\end{array}
\label{TxPDF_TotPow}%
\end{equation}
The stream $w_{k,2}$ is decoded by all cooperating users while the destination
decodes all streams.

As with previous analysis for perfectly cooperating transmitters, in
evaluating the value of a coalition we assume that the users outside a
coalition cooperate to act as worst case jammers and transmit Gaussian signals
that are independent of the signals of the users in the coalition. We show
that the \textit{PDF jamming game} is an NTU game. To this end, we first
determine the PDF rate region by applying the result in \cite[Thrm.
1]{LSNGKRNBM:01}. Let $\mathcal{G}\subseteq\mathcal{S}$ and $\mathcal{G}^{c}$
be the complement of $\mathcal{G}$ in $\mathcal{S}$. We write $R_{\mathcal{G}%
,j}$ $=$ $%
{\textstyle\sum\nolimits_{m\in\mathcal{G}}}
R_{m,j}$, $j=1,2$, $R_{\mathcal{G}}=R_{\mathcal{G},1}+R_{\mathcal{G},2}$, and
the cardinality of $\mathcal{G}$ as $\left\vert \mathcal{G}\right\vert $.

\begin{theorem}
For the PDF jamming game, a rate tuple for a coalition $\mathcal{S}$ is
achievable if, for all $\mathcal{G}\subseteq\mathcal{S}$, it satisfies%
\begin{align}
R_{\mathcal{G},2}  &  \leq\min\nolimits_{m\in\mathcal{G}^{c}}\left\{
I(V_{\mathcal{G}};Y_{m}|X_{m},U,V_{\mathcal{G}^{c}})\right\}
\label{MAC_GF_source_bounds}\\
R_{\mathcal{G},1}  &  \leq I(X_{\mathcal{G}};Y_{d}|X_{\mathcal{G}^{c}%
},V_{\mathcal{S}},U)\label{MAC_GF_dest_bound1}\\
R_{\mathcal{S}}  &  \leq I(X_{\mathcal{S}};Y_{d}). \label{MAC_GF_dest_bound2}%
\end{align}

\end{theorem}

\begin{proof}
The proof follows directly from \cite[Thrm. 1]{LSNGKRNBM:01} assuming worst
case jamming from users outside $\mathcal{S}$.
\end{proof}

A bound on the sum-rate $R_{\mathcal{G},2}$, for all $\mathcal{G}%
\subset\mathcal{S}$, results from jointly decoding the messages $w_{k,2}$, for
all $k\in\mathcal{G}$, at a cooperating user $m\not \in \mathcal{G}$. We
obtain the bound in (\ref{MAC_GF_source_bounds}) by taking the smallest bound
over all such $m\in\mathcal{S}$. The bound in (\ref{MAC_GF_dest_bound1})
results from decoding $w_{k,1}$, for all $k\in\mathcal{G}$, at the
destination. Finally, the bound in (\ref{MAC_GF_dest_bound2}) results from
decoding all messages at the destination. We obtain the bounds on
$R_{\mathcal{G}}$, for all $\mathcal{G}\subseteq\mathcal{S}$, by summing the
bounds on $R_{\mathcal{G},1}$ and $R_{\mathcal{G},2}$. The bounds on
$R_{\mathcal{G},1}$ are given by (\ref{MAC_GF_dest_bound1}). We henceforth
denote this bound as $B_{\mathcal{G},1}$. On the other hand, in addition to
the bound in (\ref{MAC_GF_source_bounds}), for any partition $(\mathcal{G}%
_{1},\mathcal{G}_{2},\ldots,\mathcal{G}_{N})$ of $\mathcal{G}$ such that
$1\leq N\leq\left\vert \mathcal{G}\right\vert $, a bound on $R_{\mathcal{G}%
,2}$ is obtained as a sum of the bounds on $R_{\mathcal{G}_{n}}$, i.e., from
the fact that $R_{\mathcal{G},2}=\sum\nolimits_{n=1}^{N}R_{\mathcal{G}_{n},2}%
$. Thus, the smallest bound on $R_{\mathcal{G},2}$ is a minimum over all such
partitions. Let $(\mathcal{G}_{1}^{\ast},\mathcal{G}_{2}^{\ast},\ldots
,\mathcal{G}_{N}^{\ast})$ be the minimizing partition. Further, from
(\ref{MAC_GF_source_bounds}), we see that for each $\mathcal{G}_{n}^{\ast}$,
there exists an index $m_{n}^{\ast}$ denoting the decoding user at which the
bound on $R_{\mathcal{G}_{n}^{\ast},2}$ is a minimum. We write this smallest
bound on $R_{\mathcal{G},2}$ as $B_{\mathcal{G},2}\left(  \left\{
\mathcal{G}_{n}^{\ast},m_{n}^{\ast}\right\}  _{N}\right)  $ to denote the
dependence of the bound on the minimizing partition and indexes such that
$R_{\mathcal{G}}\leq B_{\mathcal{G},1}+B_{\mathcal{G},2}\left(  \left\{
\mathcal{G}_{n}^{\ast},m_{n}^{\ast}\right\}  _{N}\right)  $. We obtain an
achievable rate region for the users in a coalition $\mathcal{S}$ by
substituting (\ref{TxPDF_Xk}) in (\ref{MAC_GF_source_bounds}%
)-(\ref{MAC_GF_dest_bound2}) for each choice of $(p_{k,d},p_{k,c},p_{k,u})$
subject to (\ref{TxPDF_TotPow}) and for all $k\in\mathcal{S}$. We write
$\underline{P}$ to denote the vector of tuples $(p_{k,d},p_{k,c},p_{k,u})$ for
all $k\in\mathcal{S}$ and $\mathcal{R}_{\mathcal{S}}\left(  \underline
{P}\right)  $ for the rate region achieved for each choice of $\underline{P}$.
For this signaling, the bounds in (\ref{MAC_GF_dest_bound1}) are concave
functions of $p_{k,d}$ while that in (\ref{MAC_GF_dest_bound2}) depend only
$p_{k}$ and $p_{k,u}$ for all $k\in\mathcal{G}$. However, the bounds in
(\ref{MAC_GF_source_bounds}) are not concave functions since they include
interference from $p_{k,d}$ for all $k\not =m$. Thus, the PDF rate region
$\mathcal{R}_{\mathcal{S}}^{PDF}$ is obtained as
\begin{equation}
\mathcal{R}_{\mathcal{S}}^{PDF}=co\left(  \bigcup\nolimits_{\underline{P}%
}\mathcal{R}_{\mathcal{S}}\left(  \underline{P}\right)  \right)
\label{PDF_RS}%
\end{equation}
where $co$ denotes the convex hull operation. Further, each rate tuple on the
hull may be achieved by a different $\underline{P}$. We define the value,
$\mathcal{V}(\mathcal{S})$, of a coalition $\mathcal{S}$ as a $K$-dimensional
rate region where the rates achieved by the users in $\mathcal{S}$ belongs to
the largest achievable $\mathcal{R}_{\mathcal{S}}^{PDF}$ while those for the
users not in $\mathcal{S}$ can take arbitrary values in the $\left\vert
\mathcal{S}^{c}\right\vert $-dimensional orthant $\mathbb{R}_{+}^{\left\vert
\mathcal{S}^{c}\right\vert }$. For this $\mathcal{V}(\mathcal{S})$, from
Definition \ref{NTU_def}, the PDF jamming game is an NTU game.

To determine the core of this game, one has to verify if the game is cohesive,
i.e., if the GC rate region, $\mathcal{V}(\mathcal{K})$, satisfies
(\ref{NTU_Cohesive}) for all partitions $(\mathcal{S}_{1},\mathcal{S}%
_{2},\ldots,\mathcal{S}_{N})$, $2\leq N\leq K$. We begin by determining the
power allocations that maximize the region $\mathcal{R}_{\mathcal{S}}^{PDF}$
for any coalition. Maximizing $\mathcal{R}_{\mathcal{S}}^{PDF}$ is not a
straightforward optimization problem since the rate bounds in
(\ref{MAC_GF_source_bounds}) are not in general concave functions of
\underline{$P$}. To alleviate this problem, we will build on the result in
\cite[Proposition 1]{UlukusKaya:01} where it has been shown that for a
two-user cooperative MAC, irrespective of the channel gains, the power
allocation that maximizes the rate region simplifies to setting either
$p_{k,d}=0$ or $p_{k,c}=0$ for all $k$ and for all choices of $p_{k}$ and
$p_{k,u}$ subject to (\ref{TxPDF_TotPow}). This allocation also simplifies the
rate bounds to concave functions of power that can be maximized using convex
optimization techniques. We prove a similar result for the clustered model and
for arbitrary $K$. Our result has the intuitive interpretation that the
clustered users benefit from exploiting their strong inter-user gains to
decode and forward all messages for each other, i.e., in addition to $w_{0}$,
each user transmits only one message stream which is decoded by all other
cooperating users.

\begin{theorem}
\label{Th_PDF}The rate region $R_{\mathcal{S}}^{PDF}$ of a coalition
$\mathcal{S}$ of clustered users, for all $\mathcal{S}\subseteq\mathcal{K}$,
is maximized when user $k$ sets $p_{k,d}=0$, for all $k\in\mathcal{S}$.
\end{theorem}

%

\begin{figure}
[ptb]
\begin{center}
\includegraphics[
height=1.1666in,
width=3.5535in
]%
{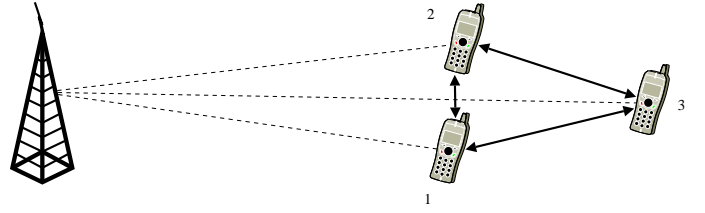}%
\caption{{}A three-user clustered MAC.}%
\label{Fig_PDFGeom}%
\end{center}
\end{figure}

\begin{proof}
We assume Gaussian signaling for all the users in $\mathcal{K}$. For the users
in $\mathcal{S}$ we choose the signals as in (\ref{TxPDF_Xk}) and fix the
transmit and cooperative powers $p_{k}$ and $p_{k,u}$ such that the remaining
power $\tilde{p}_{k}\overset{\vartriangle}{=}p_{k}-p_{k,u}$ is split between
$p_{k,d}$ and $p_{k,c}$ for all $k\in\mathcal{S}$. The region $\mathcal{R}%
_{\mathcal{S}}^{PDF}$ is given by (\ref{PDF_RS}) for all choices of
$(p_{k},p_{k,c},p_{k,u})$. We develop the results for the $\left\vert
\mathcal{S}\right\vert $-user sum-rate bound $R_{\mathcal{S}}$. One can extend
the proof in a straightforward manner to the bounds on $R_{\mathcal{G}}$ for
any $\mathcal{G}\subset\mathcal{S}$. Without loss of generality, we write the
jamming noise seen by a coalition $\mathcal{S}$ as $\left(  J_{\mathcal{S}%
}-1\right)  $ and scale the signal powers in (\ref{TxPDF_TotPow}) for all
$k\in\mathcal{S}$ by $J_{\mathcal{S}}$ such the total interference and noise
power is unity. Since the bound in (\ref{MAC_GF_dest_bound2}) is independent
of $p_{k,d}$ and $p_{k,c}$ for a fixed $p_{k}$ and $p_{k,u}$, we focus on the
bounds on $R_{\mathcal{S}}$ obtained as a sum of the bounds on $R_{\mathcal{S}%
,1}$ and $R_{\mathcal{S},2}$. Let $\left\{  \mathcal{S}_{n}^{\ast},m_{n}%
^{\ast}\right\}  _{N}$ be a partition of $\mathcal{S}$ and a collection of
indexes that jointly achieve the smallest bound on $R_{\mathcal{S},2}$ such
that
\begin{equation}
R_{\mathcal{S}}\leq B_{\mathcal{S},1}+B_{\mathcal{S},2}\left(  \left\{
S_{n}^{\ast},m_{n}^{\ast}\right\}  _{N}\right)  \label{PDF_Th_RS}%
\end{equation}
where as described earlier $B_{\mathcal{S},1}$ and $B_{\mathcal{S},2}$ are the
smallest bounds on $R_{\mathcal{S},1}$ and $R_{\mathcal{S},2}$, respectively.
For the Gaussian signaling in (\ref{TxPDF_Xk}), using (\ref{TxPDF_Yd}) and
(\ref{TxPDF_Yk}) these terms simplify as%
\begin{equation}
B_{\mathcal{S},1}=\log\left(  1+\sum\nolimits_{i\in\mathcal{S}}h_{d,i}%
p_{i,d}\right)  \label{PDF_AS}%
\end{equation}%
\begin{equation}
B_{\mathcal{S},2}\left(  \left\{  S_{n}^{\ast},m_{n}^{\ast}\right\}
_{N}\right)  =\log\prod_{n=1}^{N}\left(  1+\frac{\sum\limits_{i\in
\mathcal{S}_{n}^{\ast}}h_{m_{n}^{\ast},i}p_{i,c}}{1+\sum\limits_{j\in
\mathcal{S},j\neq m_{n}^{\ast}}h_{m_{n}^{\ast},j}p_{j,d}}\right)
.\label{PDF_LS}%
\end{equation}
Observe that $B_{\mathcal{S},1}~$in (\ref{PDF_AS}) is an increasing function
of $p_{i,d}$ while $B_{\mathcal{S},2}$ in (\ref{PDF_Th_RS}) and (\ref{PDF_LS})
is decreasing in $p_{i,d}$, for all $i\in\mathcal{S}$. Therefore, it is not
immediately clear whether setting $p_{i,d}=0$ for all $i\in\mathcal{S}$ would
maximize the bounds in (\ref{PDF_Th_RS}). Consider the case where $p_{i,d}=0$,
such that $p_{i,c}=\tilde{p}_{i}$, for all $i\in\mathcal{S}$. Denoting the
minimizing partitions and indexes for this case by $S_{n}$ and $m_{n}$,
respectively, for all $n=1,2,\ldots,N$, the bounds in (\ref{PDF_Th_RS})
simplify as
\begin{equation}
\left.  R_{\mathcal{S}}\right\vert _{{p_{i,d}=0}}\leq\left.  B_{\mathcal{S}%
,2}\left(  \left\{  S_{n},m_{n}\right\}  _{N}\right)  \right\vert
_{{p_{i,d}=0}}.\label{RS_pdzero}%
\end{equation}
On the other hand, for any $p_{i,d}>0$, we denote the minimizing partitions
and indexes by $\mathcal{S}_{t}^{\prime}$ and $m_{t}^{\prime}$ where
$t=1,\ldots,T$, and rewrite (\ref{PDF_Th_RS}) as%
\begin{equation}
\left.  R_{\mathcal{S}}\right\vert _{{p_{i,d}>0}}\leq B_{\mathcal{S}%
,1}+B_{\mathcal{S},2}\left(  \left\{  S_{t}^{\prime},m_{t}^{\prime}\right\}
_{T}\right)  .\label{RS_pd_Nzero}%
\end{equation}
Using the identity $(1+\sum_{k}x_{k})\leq\Pi_{k}(1+x_{k})$, for all $x_{k}>0$,
we upper bound $B_{\mathcal{S},1}$, and thus, $\left.  R_{\mathcal{S}%
}\right\vert _{{p_{i,d}>0}}$ in (\ref{RS_pd_Nzero}) with
\begin{align}
&  \log\left[  \left(  1+\sum_{i\in\mathcal{S}_{1}}h_{d,i}p_{i,d}\right)
\ldots\left(  1+\sum_{i\in\mathcal{S}_{N}}h_{d,i}p_{i,d}\right)  \right]
\label{PDF_RS_B1}\\
&  \text{ \ \ \ }+B_{\mathcal{S},2}\left(  \left\{  S_{t}^{\prime}%
,m_{t}^{\prime}\right\}  _{T}\right)  \nonumber\\
&  \leq\log\left[  \left(  1+\sum_{i\in\mathcal{S}_{1}}h_{d,i}p_{i,d}\right)
\ldots\left(  1+\sum_{i\in\mathcal{S}_{N}}h_{d,i}p_{i,d}\right)  \right]
\label{PDF_RS_FB}\\
&  \text{ \ \ \ }+B_{\mathcal{S},2}\left(  \left\{  S_{n},m_{n}\right\}
_{N}\right)  \nonumber
\end{align}
where the inequality in (\ref{PDF_RS_FB}) follows from the fact that for the
chosen values of $p_{i,d}>0$, for all $i\in\mathcal{S}$, the set $\left\{
S_{t}^{\prime},m_{t}^{\prime}\right\}  _{T}$ results in the smallest bound on
$R_{\mathcal{S},2}$. To show that the bound in (\ref{RS_pd_Nzero}) is smaller
than that in (\ref{RS_pdzero}), from (\ref{PDF_RS_B1}) and (\ref{PDF_RS_FB}),
it suffices to show that $\left.  B_{\mathcal{S},2}\left(  \left\{
S_{n},m_{n}\right\}  _{N}\right)  \right\vert _{{p_{i,d}=0}}$ is upper bounded
by
\begin{multline}
\log\left[  \left(  1+\sum_{i\in\mathcal{S}_{1}}h_{d,i}p_{i,d}\right)
\ldots\left(  1+\sum_{i\in\mathcal{S}_{N}}h_{d,i}p_{i,d}\right)  \right]
\label{PDF_Require}\\
+B_{\mathcal{S},2}\left(  \left\{  S_{n},m_{n}\right\}  _{N}\right)  .
\end{multline}
Expanding (\ref{PDF_Require}) using (\ref{PDF_AS}) and (\ref{PDF_LS}) and
rearranging the terms, we need to show that%
\begin{equation}
\prod\limits_{n=1}^{N}\left[  \frac{\left(  1+\sum\nolimits_{i\in
\mathcal{S}_{n}}h_{m_{n},i}\tilde{p}_{i}\right)  }{\left(  1+\frac{\sum
_{i\in\mathcal{S}_{n}}h_{m_{n},i}p_{i,c}}{1+\sum_{j\neq m_{n}}h_{m_{n}%
,j}p_{j,d}}\right)  }\right]  \geq\prod\limits_{n=1}^{N}\left(  1+\sum
_{i\in\mathcal{S}_{n}}h_{d,i}p_{i,d}\right)  .\label{PDF_Req2}%
\end{equation}
Simplifying (\ref{PDF_Req2}) further, it suffices to show that, for all
$n=1,2,\ldots,N$,%
\begin{equation}
\left(  1+\frac{\sum_{i\in\mathcal{S}_{n}}h_{m_{n},i}p_{i,c}}{1+\sum_{j\neq
m_{n}}h_{m_{n},j}p_{j,d}}\right)  \leq\frac{\left(  1+\sum\nolimits_{i\in
S_{n}}h_{m_{n},i}\tilde{p}_{i}\right)  }{\left(  1+\sum_{i\in\mathcal{S}_{n}%
}h_{d,i}p_{i,d}\right)  }.\label{PDF_Req3}%
\end{equation}
Recall that $\tilde{p}_{i}$ and $p_{i,c}$ are the powers for transmitting
$w_{k,c}$ when $p_{i,d}=0$ and $p_{i,d}\not =0$, respectively. For a fixed
$p_{i}$ and $p_{i,u}$, since $\tilde{p}_{i}>p_{i,c}$ we can expand
$h_{m_{n},i}\tilde{p}_{i}$ as $h_{m_{n},i}p_{i,c}+h_{d,i}p_{i,d}+(h_{m_{n}%
,i}-h_{d,i})p_{i,d}$, for all $i\in S_{n}$. We also expand the denominator of
the term to the left side of the inequality in (\ref{PDF_Req3}) over all
$j\in\mathcal{S}$ with $j\not =m_{n}$, where $m_{n}\in\mathcal{S}_{n}%
^{c}=\mathcal{S}\backslash\mathcal{S}_{n}$. With these two expansions
(\ref{PDF_Req3}) simplifies to requiring%
\begin{multline}
\frac{\sum_{i\in\mathcal{S}_{n}}h_{m_{n},i}p_{i,c}}{1+\sum_{j\in
\mathcal{S}_{n}}h_{m_{n},j}p_{j,d}+\sum_{j\in S_{n}^{c},j\neq m_{n}}%
h_{m_{n},j}p_{j,d}}\leq\label{PDFreq4}\\
\frac{\left(  \sum\nolimits_{i\in S_{n}}h_{m_{n},i}p_{i,c}+\sum\nolimits_{i\in
S_{n}}(h_{m_{n},i}-h_{d,i})p_{i,d}\right)  }{\left(  1+\sum_{i\in
\mathcal{S}_{n}}h_{d,i}p_{i,d}\right)  }.
\end{multline}
Comparing the numerators and denominators on both sides of (\ref{PDFreq4}),
for the clustered model where $h_{m_{n},i}>h_{d,i}$ for all $i$, one can
easily see that the inequality is satisfied. Thus, for any $(p_{i},p_{i,u})$,
setting $p_{i,d}=0$, for all $i$, maximizes the bound on $R_{\mathcal{S}}$.
One can similarly show that the rate bounds for all $\mathcal{G}%
\subset\mathcal{S}$ are also maximized, and thus, the region $\mathcal{R}%
_{\mathcal{S}}\left(  \underline{P}\right)  $ is maximized. Since the argument
holds for all $\underline{P}$ the rate tuples on the hull of $\mathcal{R}%
_{\mathcal{S}}^{PDF}$ are also maximized.\ 
\end{proof}

%

\begin{figure}
[ptb]
\begin{center}
\includegraphics[
height=3.2162in,
width=3.3857in
]%
{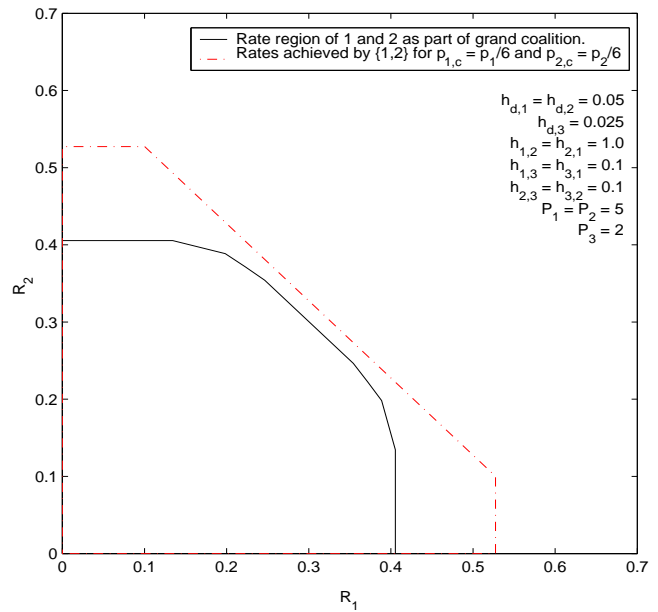}%
\caption{{}Rate regions for the GC and the $\{1,2\}$ coalition in the $R_{1}%
$-$R_{2}$ plane.}%
\label{Fig_PDF_Rateregions}%
\end{center}
\end{figure}

Thus, from Theorem \ref{Th_PDF}, setting $p_{k,d}=0$ for all $k$ simplifies
the bounds in (\ref{PDF_Th_RS})-(\ref{PDF_LS}) to concave functions of $P_{k}$
for all $k\in\mathcal{S}$. As a result, the rate region $\mathcal{R}%
_{\mathcal{S}}^{PDF}$ does not require the convex hull operation in
(\ref{PDF_RS}) thus simplifying the evaluation of $\mathcal{V}(S)$ for any
$\mathcal{S}$. From Definitions \ref{NTU_def} and \ref{NTU_Coh}, a necessary
condition for the game to be cohesive is that, for every $\mathcal{S}%
\subset\mathcal{K}$, the projection of $\mathcal{V}(\mathcal{S})$ in the rate
space of $\mathcal{S}$, i.e. $\mathcal{R}_{\mathcal{S}}^{PDF}$, is a subset of
the projection to the same space of the GC\ value set, $\mathcal{V}%
(\mathcal{K})$. While Theorem \ref{Th_PDF} allows computing $\mathcal{V}(S)$
relatively easily, in general, inferences on the cohesiveness of the game
cannot be drawn easily for arbitrary values of channel gains, user powers, and
for any $K$. We thus use an example to illustrate that the PDF\ user
cooperation game may not be cohesive, i.e., the GC may not achieve the largest
rate region. In fact, our example reveals that for asymmetric inter-user
channel gains and a few weak jammers, users can form smaller coalitions to
achieve larger rates relative to the GC.

\begin{remark}
Verifying whether the grand coalition is cohesive is equivalent to verifying
whether the conditions in (\ref{NTU_Cohesive}) hold, i.e., (\ref{NTU_Cohesive}%
) captures the functional dependence between the channel parameters and
transmit powers required for the NTU game to be cohesive. In general, however,
verifying the requirement that the intersection of the $K$-dimensional rate
regions corresponding to all possible coalition structures lies within the GC
rate region in (\ref{NTU_Cohesive}) is not straightforward. Furthermore, the
verification complexity grows exponentially in $K$.
\end{remark}

\begin{example}
Consider a cooperative MAC shown in Fig. \ref{Fig_PDFGeom} with $3$ users
labeled $1,2,$ and $3$ that are clustered as in (\ref{Cluster_Model}) with
gains $h_{d,1}=h_{d,2}=0.05$, $h_{d,3}=0.025,$ $h_{1,2}=h_{2,1}=1$,
$h_{1,3}=h_{3,1}=h_{2,3}=h_{3,2}=0.1$, and power constraints $P_{1}=P_{2}=5,$
and $P_{3}=2\,$. Thus, users~$1$ and $2$ have a stronger inter-user channel to
each other than to user $3$ while user $3$ has a smaller transmit power. In
Fig. \ref{Fig_PDF_Rateregions}, we plot the rate region achieved by $1$ and
$2$ when they are part of the grand coalition, i.e., we plot the projection of
the GC\ region $\mathcal{V}(\left\{  1,2,3\right\}  )$ on the $R_{1}$-$R_{2}$
plane computed using the bounds in (\ref{MAC_GF_source_bounds}%
)-(\ref{MAC_GF_dest_bound1}) and Theorem \ref{Th_PDF}. Also shown is the rate
region achieved by users $1$ and $2$ as a coalition $\{1,2\}$ from Theorem
\ref{Th_PDF} for $p_{1,c}=p_{2,c}=P_{1}/6$ and assuming maximum jamming by
user $3$. Since the latter region contains the former, the game is not
cohesive. Further, for every rate tuple achieved by users $1$ and $2$ when
they are a part of the GC, there exists at least a tuple where both users
achieve larger rates for the coalition $\left\{  1,2\right\}  $, and thus, the
GC is not stable. This is because the requirement of decoding the messages
from users $1$ and $2$ at the relatively distant user $3$ for the GC results
in tighter bounds than those achieved by the coalition $\left\{  1,2\right\}
$ in the presence of a weak jammer $3$.
\end{example}

\section{\label{Sec_6}Concluding Remarks}

We have studied the stability of the GC when users in a wireless network are
allowed to cooperate while maximizing their own rates. For an IC, we have
shown that when only receivers are allowed to cooperate by jointly decoding
their received signals, the GC is both stable and sum-rate optimal. However,
we have shown that if the receivers cooperated using linear multiuser
detectors, they cannot arbitrarily share the gains from cooperation and the
stability of the GC depends on the SNR regime and the detector. We have also
studied transmitter cooperation in an IC with perfectly cooperating receivers.
We have shown that when transmitters are allowed to cooperate via noise-free
links the GC is sum-rate optimal but may not be stable. Finally, we have shown
that for a network where clustered transmitters cooperate by mutually decoding
messages via PDF, the optimality of the GC from both a stability and a rate
region perspective depends on the network geometry and the jamming potential
of the users. For transmitter cooperation, we have presented a jamming
interpretation to characterize the value of a coalition. Although the
assumption represents an extreme adversarial response of the complementary
coalition, it lower bounds the rates achieved by a coalition that breaks away
from the GC, and is therefore a strong result when the core is empty. Our work
has also demonstrated that stability depends on the incentives and
disincentives that users have to cooperate. For example, the noise enhancement
in a decorrelating detector can act as a disincentive to the stability of the
GC in the low SNR\ regime. Similarly, channel gains and weak jammers can
destabilize the GC when transmitters cooperate perfectly. Furthermore, noisy
inter-user channels can also affect the stability of the GC for decoding transmitters.

\bibliographystyle{IEEEtran}
\bibliography{main_JSAC08}

\end{document}